\documentclass{aa}  

\usepackage{graphicx}
\usepackage{multirow}
\usepackage{booktabs}
\usepackage{adjustbox}
\usepackage{hyperref}
\usepackage{dcolumn}
\usepackage[final]{pdfpages}
\usepackage{natbib}
\usepackage{indentfirst}
\usepackage{threeparttable, tablefootnote}
\usepackage{txfonts}

\begin{document}

   \title{DEATHSTAR: Nearby AGB stars with the Atacama Compact Array 
}
   \subtitle{II. CO envelope sizes and asymmetries: The S-type stars \thanks{The reduced data cubes are only available in electronic form at the CDS via anonymous ftp to cdsarc.u-strasbg.fr (130.79.128.5)
or via \url{http://cdsweb.u-strasbg.fr/cgi-bin/qcat?J/A+A/}}}

   \author{M. Andriantsaralaza
          \inst{1},
          S. Ramstedt\inst{1}
          \and
          W. H. T. Vlemmings\inst{2}
          \and
          T. Danilovich\inst{3}
          \and
          E. De Beck\inst{2}
          \and
          M. A. T. Groenewegen\inst{4}
          \and
          S. H\"ofner\inst{1}
          \and
          F. Kerschbaum\inst{5}
          \and
          T. Khouri\inst{2}
          \and
          M. Lindqvist\inst{2}
          \and
          M. Maercker\inst{2}
          \and
          H. Olofsson\inst{2}
          \and
          G. Quintana-Lacaci\inst{6}
          \and
          M. Saberi\inst{7,8}
          \and
          R. Sahai\inst{9}
          \and
          A. Zijlstra\inst{10}
          }

   \institute{Theoretical Astrophysics, Division for Astronomy and Space Physics, Department of Physics and Astronomy, Uppsala University, Box 516, SE-751 20 Uppsala, Sweden\\
              \email{miora.andriantsaralaza@physics.uu.se}
        \and
        Department of Space, Earth and Environment, Chalmers University of Technology, Onsala Space Observatory, 439 92 Onsala, Sweden
        \and
        Department of Physics and Astronomy, Institute of Astronomy, KU Leuven, Celestijnenlaan 200D, 3001 Leuven, Belgium
        \and
        Koninklijke Sterrenwacht van Belgi\"e, Ringlaan 3, 1180 Brussels, Belgium
        \and
        Department of Astrophysics, University of Vienna, T\"urkenschanzstr. 17, 1180 Vienna, Austria
        \and
        Group of Molecular Astrophysics. IFF. CSIC. C/ Serrano 123, E-28006, Madrid, Spain
        \and
        Rosseland Centre for Solar Physics, University of Oslo, P.O. Box 1029 Blindern, NO-0315 Oslo, Norway
        \and
        Institute of Theoretical Astrophysics, University of Oslo, P.O. Box 1029 Blindern, NO-0315 Oslo, Norway
        \and
        Jet Propulsion Laboratory, MS 183-900, California Institute of Technology, Pasadena, CA 91109, USA
        \and
        Jodrell Bank Centre for Astrophysics, Alan Turing Building, University of Manchester, Manchester M13 9PL, UK
        }

   \date{Received March 31, 2021; accepted June 3, 2021}

 
  \abstract
   {}
   {We aim to constrain the sizes of, and investigate deviations from spherical symmetry in, the CO circumstellar envelopes (CSEs) of 16 S-type stars, along with an additional  7 and 4 CSEs of C-type and M-type AGB stars, respectively.  
   }
   {We map the emission from the CO $J=\,$2--1 and 3--2 lines observed with the Atacama Compact Array (ACA) and its total power (TP) antennas, and fit with a Gaussian distribution in the uv- and image planes for ACA-only and TP observations, respectively. The major axis of the fitted Gaussian  for the CO(2--1) line data gives a first estimate of the size of the CO-line-emitting CSE. We investigate possible signs of deviation from spherical symmetry by analysing the line profiles and the minor-to-major axis ratio obtained from visibility fitting, and by investigating the deconvolved images.}
   {The sizes of the CO-line-emitting CSEs of low-mass-loss-rate (low-MLR) S-type stars fall between the sizes of the CSEs of C-stars, which are larger, and those of M-stars, which are smaller, as expected because of the differences in their respective CO abundances and the dependence of the photodissociation rate on this quantity. The sizes of the low-MLR S-type stars show no dependence on circumstellar density, as measured by the ratio of the MLR to terminal outflow velocity, irrespective of variability type.  The density dependence steepens for S-stars with higher MLRs. While the CO(2--1) brightness distribution size of the low-density S-stars is in general smaller than the predicted photodissociation radius (assuming the standard interstellar radiation field), the measured size of a few of the high-density sources is of the same order as the expected photodissociation radius. Furthermore, our results show that the CO CSEs of most of the S-stars in our sample are consistent with a spherically symmetric and smooth outflow. For some of the sources, clear and prominent asymmetric features are observed which are indicative of intrinsic circumstellar anisotropy.
   }
   {As the majority of the S-type CSEs of the stars in our sample are consistent with a spherical geometry, the CO envelope sizes obtained in this paper will be used to constrain detailed radiative transfer modelling to directly determine more accurate MLR estimates for the stars in our sample. For several of our sources that present signs of deviation from spherical symmetry, further high-resolution observations would be necessary to investigate the nature of, and the physical processes behind, these asymmetrical structures. This will provide further insight into the mass-loss process and its related chemistry in S-type AGB stars. }

\titlerunning{DEATHSTAR: Nearby AGB stars with the ACA: The S-stars}
\authorrunning{M. Andriantsaralaza et al.}
   \maketitle
%
\section{Introduction}
The evolutionary path of a star mainly depends on its initial mass. Low-to-intermediate mass stars ($\sim0.8<M<8\,\mathrm{M}_\odot$) evolve into asymptotic giant branch (AGB) stars near the end of their lives. Although stellar winds are common phenomena in stars, AGB stars are subject to slow and massive winds, with mass-loss rates (MLRs) ranging from $10^{-8}$  to as high as $10^{-4} \,\mathrm{M_\odot\,yr}^{-1}$ \citep[e.g.][]{olofsson_1999,watcher2002,Hofner2018}. A more recent study suggests an upper limit on the MLR of about a few times $10^{-5}\,\mathrm{M_\odot\,yr}^{-1}$ \citep{Decin2019}, in better agreement with current wind models \citep{Eriksson2014, Bladh2019}. During the AGB phase, the mass loss of the star determines its evolution \citep{bloecker1995}. It is well-established that mass loss in AGB stars is caused by radiation pressure on dust grains, pushing the grains and the surrounding gas out of the stellar gravitational field because of gas and dust momentum exchange (e.g. \citealt{susanne2015} and references therein). The material ejected through the wind creates a chemically rich expanding envelope around the AGB star, namely the circumstellar envelope (CSE) \citep{Habing1996}.  Changes in the MLR can affect the evolution of the star and its nucleosynthesis \citep{Forestini1997}. Investigating the mass-loss phenomenon is crucial in gaining a better understanding of late stellar evolution, as well as of the galactic chemical evolution, as AGB stars contribute significantly to the enrichment of the interstellar medium \citep{Matsuura2009}. 

The most reliable method to estimate MLRs is through radiative transfer modelling of CO rotational line emission combined with observations of these lines towards the CSE. Further, as opposed to the observations of dust continuum emission, these observations give a direct measure of the wind velocity \citep{schoier2001,Sofia2008}. CO is the second most abundant molecule in CSEs and the most abundant molecule with a permanent dipole moment. Its emission is directly linked to the temperature and density throughout the CSE, and its excitation properties are relatively well understood \citep[e.g.][]{Olofsson2005,Elvire2010,Saberi2019}. This MLR-determination method consists of solving the non-local thermodynamic equilibrium (NLTE) CO excitation and radiative transfer, and fitting the observed CO lines by varying the MLR and parameters related to the gas temperature distribution. It has been used in several studies \citep[e.g.][]{schoier2001,Olofsson2002,Decin2006,Sofia2006,Sofia2009,Elvire2010,Cernicharo2015} and has proven to be most successful for stars with MLRs lower than $10^{-5}$ \(\textup{M}_\odot\) yr$^{-1}$, above which additional uncertainties due to CO line emission saturation and unknown dust properties  need to be taken into consideration \citep{schoier2001, Sofia2008}. 

An accurate determination of the MLR ideally requires combination of spatially resolved observations of several CO transitions. Different CO lines are sensitive to different parameters in the model. As different transitions probe slightly different regions in the CSE depending on their excitation requirements, the MLR obtained from a range of different transitions probes a larger part of the CSE. In this paper, we study the CO(2--1)-emitting region, which overlaps with that of CO(1--0) to a large extent, and probes the cooler gas in the outer regions, and is thus closely representative of the entire size of the CO-emitting region of the CSE. In addition, the combination of several lines permits the determination of the gas temperature distribution. The MLR derived from CO observations is the average MLR that created the CSE probed by the lines. A major uncertainty with this method for determining the MLR is the poorly constrained size of the CO envelope. \citet{sunburn2015} discusses the dependence of the outer CO shell radii around AGB stars on metallicity and stellar density. The most recent size estimate of the CO envelope is based on a photodissociation model by \citet{Saberi2019}, where a standard interstellar radiation field is assumed that was developed from a previous photodissociation model by \citet{Mamon1988}. Based on an improvement in calculations of the depth dependency of the CO photodissociation rate using a line dissociation method, the most updated high-resolution CO spectroscopic data, and a larger parameter set, the results of \citet{Saberi2019} show that CO envelope sizes were systematically overestimated by 1--40 percent for a significant number of C-type stars. This overestimation results in an uncertainty of the same order in the MLR to a first approximation. \\

Constraining the size of the CO-emitting envelope is the first step of the DEATHSTAR\footnote{\url{www.astro.uu.se/deathstar}} project, the overall aim of which is to provide more accurate MLR estimates by directly measuring the CO(2--1) line-emitting envelope sizes. The first results of the project are presented in \citet{deathstar}, where the CO-envelope properties of a sample of 42 C-type and M-type stars were analysed. In the present paper, we aim to constrain the size of the CO CSEs of the southern S-stars of the DEATHSTAR sample. The sample selection and the completeness of the full DEATHSTAR sample are discussed in \citet{deathstar}. 

The size determination that we present in this paper gives additional important constraints on the wind properties of S-type stars. S-type stars exhibit ZrO bands that are traditionally thought to be indicative of a C/O ratio close to unity in the atmosphere, making them possible transition objects between M- and C-type stars. However, \citet{VanEck2017} showed that S-star spectra can be compatible with C/O ratios as low as 0.5, similar to those of M-type stars, using stellar atmosphere models. In addition, a study by \citet{Sofia2006, Sofia2009} indicated that the winds of the S-, M-, and C-type stars are driven by the same mechanism, but other studies, for example  \citet{schoier2013}, showed that the three chemical types present significantly different CSE chemical properties.  \\

This paper is organised as follows. We present the sample in Sect. 2. The observation, data reduction, and analysis are presented in Sect. 3. Section 4 outlines our results, where we discuss the measured sizes and the size--density correlation for our sources, investigate the possible indications of asymmetry, and present the detection of molecules other than $^{12}$CO. Section 5 closes the paper with a discussion and a summary.
\section{The sample}
The sample of sources covered in this paper consists of 16 S-type AGB stars.
We also report observations of 7 C-type and 4 M-type stars, late additions to the DEATHSTAR sample, for which data were acquired after the completion of the study by \citet{deathstar}. All sources are listed in Table \ref{tab:sample} with their variability type as in the General Catalogue of Variable Stars (GCVS; \citealt{GCVS}), their wind properties (MLR, velocity expansion), and their distance derived by \citet{Sofia2009}, along with their distances  from the early third  Gaia data release (Gaia eDR3; \citealt{GaiaDR3}). A preliminary discussion on the reliability of the Gaia distances for the C- and M-stars in the DEATHSTAR sample is presented in \citet{deathstar}. The different distance estimates for the complete sample will be discussed in a future publication (Andriantsaralaza et al. 2021, in prep.).  

The S-type AGB stars in our sample are southern sources, with declinations lower than $+15^{\circ}$, presented in \citet{Sofia2009}. They are intrinsic S-stars in the thermal pulse phase of AGB evolution as they present Tc lines and infrared excess, have good-quality flux measurements in the 12, 25, and 60 $\mu$m bands in the IRAS Point Source Catalogue, and are part of the General Catalogue of Galactic S-stars \citep{catalogueS, jorissen1998}. 
The sources cover a large range in MLRs and expansion velocities.  The MLRs vary from $4 \times 10^{-8}$ to $3 \times 10^{-6}$ M$_\odot$ yr$^{-1}$, and the expansion velocity ranges from 2.8 to 17.2$\,$km s$^{-1}$ for the S-type stars \citep{Sofia2009}. 

The C- and M-type stars presented in this paper add up to the sample of 42 southern C- and M-type AGB stars presented in the first DEATHSTAR publication \citep{deathstar}. They represent a range of MLRs that reach up to $\sim 1 \times 10^{-5}$ M$_\odot$ yr$^{-1}$ from previous estimates, and an expansion velocity varying from 4.5 to 16 km s$^{-1}$ \citep{schoier2001,Gonzalez2003,Danilovich2015}. All sources in our sample were previously detected in CO using single-dish observations \citep{Sofia2006,Sofia2009}.   
\section{Observation, data reduction, and analysis}
\subsection{Observation with the Atacama Compact Array}
Interferometric observations of the sample sources were carried out with the Atacama Compact Array (ACA) in stand-alone mode in Cycle 5 in Bands 6 and 7, with the exception of AQ Sgr which was previously observed during Cycle 4 and published in \citet{deathstar}, but re-observed in Cycle 5 in Band 7 only. The  observation setups are identical to those of the observations presented in \citet{deathstar}. Each band was comprised of four science spectral windows that were centred on 216.4, 218.3, 230.7, and 232.1 GHz in Band 6; and centred on 330.75, 332.25, 343.52, and 345.6 GHz in Band 7. The observations covered a number of molecules including carbon monoxide emission lines: $^{12}$CO $J=2-1$, $3-2$ and $^{13}$CO $J=3-2$, as well as other molecules such as SiO, SiS, CS, and SO$_2$. The imaged data have a spectral resolution of 0.75 km s$^{-1}$ for the carbon monoxide spectral windows in both bands. The other spectral windows were set to a spectral resolution of $1.5$ and $1$ km s$^{-1}$ in Band 6 and 7, respectively, to improve sensitivity. The maximum recoverable scale with the 7-m ACA dishes at the desired science frequencies was on average $\sim 25\arcsec$ and $\sim 19\arcsec$ in Band 6 and 7, respectively.

The data were calibrated using the standard scripts provided by the pipeline using the Common Astronomy Software Application package \citep[\textsc{casa};][]{casa}. All spectral windows in both bands were self-calibrated using two channels across the peak $^{12}$CO emission for all sources to improve the signal-to-noise ratio (S/N). This increased the strength of the signal by 10 percent on average. The data were re-imaged with the \textsc{casa} task $tclean$ using the \textit{H\"ogbom} algorithm, a Briggs weighting (robust parameter = $0.5$), and $10\,000$ iterations. Sources brighter than $20$ Jy were cleaned interactively by visually checking the noise level and manually selecting the shape of the masks applied on the emission regions in each channel map before each cleaning cycle. 
\subsection{Observation with total power and data combination}
Four of the additional C- and M-type stars observed in Cycle 5, AI Vol, AFGL 3068, R Dor, and IRC-30398 (marked with * in Table 1) were also observed with the total power (TP) antennas to recover missing flux due to their large angular scales. All four sources were scanned to map the CO(2--1) emission using four 12-m TP antennas, with the exception of  AFGL$\,$3068 which was observed with three antennas, with a total time on source of 36$\,$min$\,$24 s each. The data was retrieved from the ALMA archive and calibrated using the standard \textsc{casa} procedures. The calibrated data were exported into \textsc{casa} images  using the \textsc{casa} task \textit{sdimaging} and reframed to match the rest frequencies and channel numbers of the ACA data. This was done one by one for each antenna for a given source, giving an image for each antenna and its corresponding weight as outputs. The images were then combined into one single weighted image for each source. The size of the beams of the generated TP images is  $\sim 28 \,\mathrm{x}\, 28 \, \mathrm{arcsec}$ for all four sources. The combined TP images were used as input to generate the corresponding visibilities using the $tp2vis$ algorithm. The generated measurement sets were imaged together with the ACA data using the \textsc{casa} task $tclean$. We noticed that cleaning using the \textit{H\"ogbom} algorithm led to negative bowls around the sources. In addition, we tried several sets  of iterations (from 10 000 to 100 000) and did not see any signs of artefacts introduced in the images. By checking the residual images, however, we noticed that flux was lost when applying too few iterations. Therefore, we run $tclean$ with the $multiscale$ algorithm and $100\,000$ iterations for all four sources. After data combination, the flux of the four observed sources increased by at least $30$ percent.\\

To complete our analysis, we include CO(3--2) data of W$\,$Aql in Band 7 observed with the ACA and previously published by \citet{Sofia2017}. The retrieved data were reduced and imaged using standard \textsc{casa} procedures. The resulting synthesised beam, the position angle, and the rms noise levels measured in the emission-free channels for all sources for the $^{12}$CO lines in both bands are listed in Table \ref{tab:imaging_results}.
\subsection{Fitting the emission distribution}
We fit a Gaussian emission distribution model  to the data visibilities in CO(2--1) and CO(3--2) in each velocity channel for the sources observed with ACA-only using the \textsc{uvmultifit} library implemented in  \textsc{casa} \citep{uvmultifit}. Such fitting of the  emission distribution permits initial estimates of the sizes of the CO envelopes and will reveal signs of deviation from spherical symmetry. For each run, \textsc{uvmultifit} gives the position, flux, major axis, ratio between minor and major axes, and the position angle of the fitted Gaussian.  This method takes away any bias and artefacts introduced by processes such as gridding and weighting in deconvolved images, and permits the estimation of the size of small sources. 

The emission distributions of AI$\,\,$Vol, AFGL$\,\,$3068, R$\,\,$Dor, and IRC-30398 were fitted by a Gaussian model in the image plane on the CO(2--1) TP-only data. The resulting deconvolved major axes give an estimate of the diameters of the CO envelopes. 
\section{Results}
\subsection{Line profiles}
The spectra of the $^{12}$CO $J=2-1$ and $3-2$ lines are presented in Figs \ref{fig:lines_profiles_S1}-\ref{fig:lines_profiles_S2} for all sources. The lines were generated using an aperture of 20$\arcsec$ for 2--1 and $15\arcsec$ for 3--2 set by the maximum recoverable scale of the ACA observations. For a few sources, a larger aperture was used ($<25\arcsec$) to ensure that the full extent of the emitting region was covered.  The peak flux ($F_\mathrm{peak}$), the central velocity ($\upsilon_\mathrm{c}$), and the velocity widths ($\Delta \upsilon$) of each line are listed in Table \ref{tab:imaging_results}. The peak flux is defined as the maximum point across the line. The centre velocity is the average of the two points at the extreme velocities that correspond to five-percent of the peak flux. The velocity width between the two five-percent-peak-flux points gives the total width. Some line profiles show interesting features and shapes such as wide bases (e.g. Z$\,\,$Ant) and triangular shapes (e.g. FU$\,\,$Mon). Despite the high S/N reached for most of the sources, the profiles of the weakest-emission stars (e.g. UY$\,\,$Cen, and V996$\,\,$Cen) are relatively noisy (mean S/N$<7$).
 \subsection{Gaussian distribution fitting}
Table 2 lists the angular photodissociation diameter of the sources, given by $2({R_\mathrm{p}}/{D})$, where $D$ is the distance to the sources listed in column 5 of Table \ref{tab:sample}, and $R_\mathrm{p}$ the photodissociation radius. The photodissociation region corresponds to the region where  molecules such as CO are destroyed by the interstellar radiation field. This represents the physical boundary of the CO-rich CSE of the AGB star with respect to the interstellar medium, and  is therefore a measure of the radius of the CO-CSE shell.  By definition, $R_\mathrm{p}$ corresponds to the radius where the CO abundance has dropped to half of its initial value. The photodissociation radius $R_\mathrm{p}$ is calculated using the source parameters in Table \ref{tab:sample} with Eqns 10 and 11 from \citet{schoier2001} based on the CO photodissociation model by \citet{Mamon1988} and a CO abundance, $f_0$, of $6\times10^{-4}$ for the S-stars, $2\times10^{-4}$ for the M-stars, and $10^{-3}$ for the C-stars. 
The measured major axes and ratio of the minor to major axis of the best-fitting Gaussian at the central channel (see Sect.$\,$3.3) for the CO lines are presented in Table 2 to give a first indication of the symmetry or sphericity of the sources. We take the error as the average error from the two channels on each side of the central velocity channel. Very large errors are indicated by a colon (:).\\ 

The CO(2--1) Gaussian major axes of the semi-regular and irregular S-type stars are in general about two times smaller than their angular photodissociation diameter. The case of Mira stars, however, is less conclusive because of missing flux, represented by spikes in the fitted intensity and size distributions as seen in Fig. \ref{fig:UV_fit_S1}-\ref{fig:UV_fit_S2}. Negative regions emerge, which is due to the flux being resolved out. As a consequence, the emission cannot be properly fitted by a Gaussian distribution, leading to fluctuating spikes. When the spikes are located across the central velocity channels, which should cover the largest emission region, the value of the corresponding major axis dramatically decreases. Therefore, the sizes estimated in such cases are not representative of the general behaviour of the size distribution of the star (e.g. ST$\,\,$Sgr). 

We look at the minor-to-major axis ratio obtained from the visibility fitting to get a first estimate of how circular our sources are, as size estimates are more reliable for circular sources \citep[see][]{deathstar}. The majority of the S-type stars (8 out of 15) and  almost all of the additional C- and M-type stars observed with the ACA-only in our sample have an axis ratio close to 1, within 10 percent. This indicates that most of the sources in our sample are more or less circular. 

Figure \ref{fig:size-density} shows the size of CO(2--1) CSEs obtained from the major axes of the Gaussian fits at the central velocity (in AU) plotted against $\Dot{M}/\upsilon_\infty$, ($\mathrm{M}_\odot \,$s$\,$ km$^{-1}$yr$^{-1}$), a proxy of the density, for the 25 sources discussed in this paper, along with the 42 C- and M-stars in Ramstedt et al (2020). The uncertainties in Fig. \ref{fig:size-density} are mainly caused by uncertainties on the distance \citep{deathstar}. The photodissociation diameters from \citet{Saberi2019} and \citet{Mamon1988} are represented by the dashed and thin dotted lines, respectively. The solid lines show a spline-fit to the Gaussian full width at half maximum (FWHM) of the CO(2--1) line obtained after radiative transfer modelling of the models from \citet{Saberi2019}, thus representing the expected CO(2--1) diameters.

The different trends for the sizes of low- and high-MLR stars are evident in Fig. \ref{fig:size-density}. Low-density S-stars ($<5\times10^{-8}\mathrm{M}_\odot \,$s$\,$ km$^{-1}$yr$^{-1}$) present a weak dependence on density and are more or less scattered below 2$\,$000 AU. The measured diameters are in general smaller than the expected CO(2--1) diameters based on photodissociation models (solid lines in Fig. \ref{fig:size-density}) for those with the lowest MLRs.   S-stars with a higher MLR present a steeper density dependence. 
The additional C- and M-type stars follow a similar trend: those with low density ($<5\times10^{-8}\mathrm{M}_\odot \,$s$\,$ km$^{-1}$yr$^{-1}$) show very little density dependence, while the dense ones follow a steeper correlation.  \\
\subsection{Image fitting}
\begin{figure}
   \centering
   \includegraphics[scale=0.33]{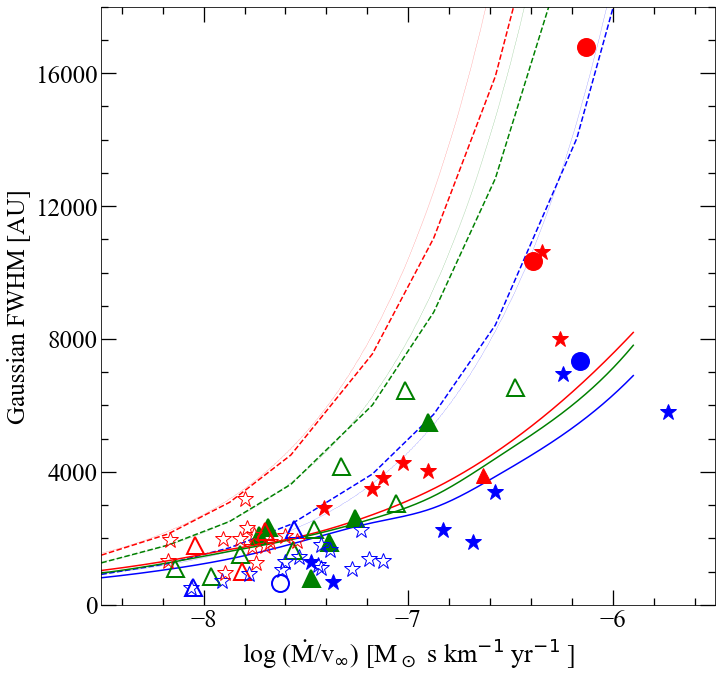}
   \caption{Major axis of the best-fitting Gaussian at the central velocity channel of the CO (2--1) emission as a function of a proxy of circumstellar density. Here, C-, S-, and M-type stars are red, green, and blue, respectively. The stars discussed in this paper are represented by triangles (ACA-only) and round (TP) symbols, while the star symbols show the data published in \citet{deathstar}. Mira-type variables are marked with solid symbols and other variables are with open symbols. The dashed lines show the photodissociation diameter of C-type (red), S-type (green), and M-type (blue) stars from \citet{Saberi2019},  while the thin lines are the fits to the results from  \citet{Mamon1988}. The solid lines show a spline-fit to the expected Gaussian FWHM of the CO(2--1) line determined from radiative transfer modelling of the models from \citet{Saberi2019}.}
              \label{fig:size-density}%
\end{figure}

\begin{figure}
   \centering
   \includegraphics[scale=0.38]{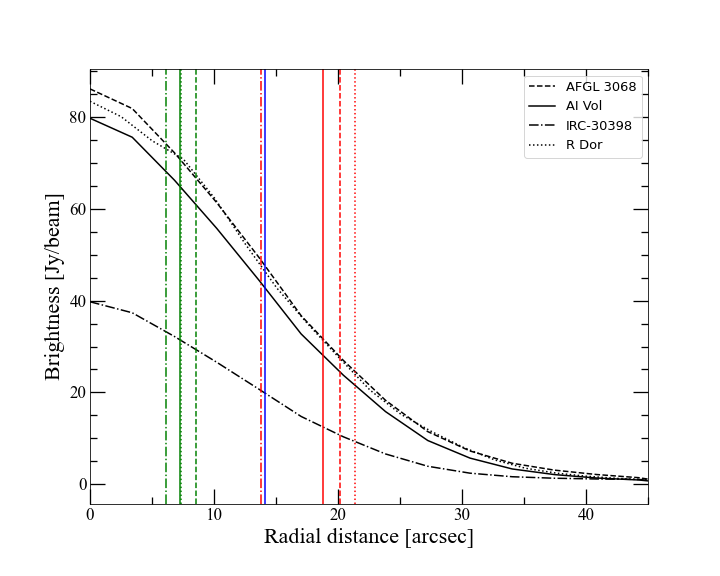}
   \caption{Observed CO(2--1) intensity distribution as a function of radius for the four stars observed with the TP. For each star, the vertical red line represents the photodissociation radius from \citet{Mamon1988} divided by the distance, $R_\mathrm{p}/D$, and the green line indicates the beam-deconvolved FWHM radius of the Gaussian fitted in the image plane. The blue solid line represents the FWHM radius of the Gaussian beams of the TP images of the four sources.}
              \label{fig:radial-profile}
\end{figure}

It can be seen in Fig. \ref{fig:size-density} that the beam-deconvolved measured CO(2--1) sizes of the high-density CSEs obtained from fitting the TP-only data in the image plane (round symbols) are significantly larger than the expected CO(2--1)-emitting region. This is particularly true for the C-type CSEs. Figure \ref{fig:radial-profile} shows the observed $\textrm{CO(2--1)}$ brightness distribution as a function of radial distance obtained by running the \textsc{casa} task  \textsc{casairing} on the TP-only image cubes of the four stars AFGL$\,\,$3068, AI$\,\,$Vol, IRC-30398, and R$\,\,$Dor.  For all four stars, the resulting emission distribution follows a Gaussian distribution  with FWHM radii that are about twice larger than the beam-deconvolved CO(2--1) FWHM radii of the Gaussians fitted in the image plane presented in Fig. \ref{fig:size-density} and Table 2. As our TP observations cover the full extent of the CO(2--1) emission, it is of interest to compare the derived beam-deconvolved CO(2--1) FWHM diameter with the photodissociation diameter to get an estimate of the extent of the CO emission in the CSE, as the CO(2--1) line is not necessarily excited throughout the entire photodissociation envelope. The FWHM radii measured in the image plane are still a factor of
$\geq 2$ smaller than the angular photodissociation radii calculated using Eqs. 10 and 11 from \citet{schoier2001} based on the models of \citet{Mamon1988}, represented by the red lines. The photodissociation radii plotted in Fig. \ref{fig:radial-profile} are angular sizes and are  therefore affected by the uncertainties on the distances.

Using observations with the ACA only, the CO(2--1) sizes of the CSEs of AFGL$\,\,$3068 and AI$\,\,$Vol obtained by measuring the CO(2--1) FWHM from the visibility fitting are underestimated by a factor of two compared to using the TP data. Further, the effects of resolved-out flux on the ACA-only visibility fits of IRC-30398 and R$\,\,$Dor do not permit a reliable size measurement for these sources.

\citet{deathstar} found that the observed sizes of the C-type stars are larger than expected in CO(2--1) based on photodissociation theory, while the M-type stars are generally smaller. This agrees well with our results for low-density stars, but our new TP data show that C- and M-stars tend to be larger than expected at high density. We observe that, for the same density, C-stars are systematically larger than M-type stars. This is in agreement with the results of \citet{deathstar}. Looking into the spread in the size for  each chemical type for all the stars presented in Fig. \ref{fig:size-density}, we observe that C-type stars display the largest scatter, followed by M-type stars, while S-type stars are relatively clustered, which may be the result of the number of sources in each category. We investigate the deviation of the data points from the expected CO(2--1) diameters (solid lines in Fig. \ref{fig:size-density}) by calculating their mean $\chi^2$. We find that, on average, the derived sizes of the M-type stars differ the most from the expected sizes from theory, with a mean deviation that is about twice larger than that of the S- and C-stars. \citet{deathstar} investigated the reliability of a Gaussian fitting applied to the CO (2-1) line for the determination of the sizes of the CSEs of AGB stars and their photodissociation radii.  This latter study shows that intermediate- to high-MLR stars are more likely to exhibit a non-Gaussian intensity distribution. Therefore, the Gaussian major axes obtained with this method are less reliable for high-density sources ($>5\times10
^{-8}\,\mathrm{M}_\odot \,$s$\,$ km$^{-1}$yr$^{-1}$) than for the low-density cases. 

 \subsection{Asymmetrical features}
We look for asymmetrical features in the line profiles and in the results of the Gaussian fits of the observed sources. The asymmetries are divided into two categories depending on whether they are consistent with a spherically symmetric CSE or not.
\subsubsection{Features consistent with a smooth flow}
A number of the S-sources in our sample suffer from anomalies linked to how the observations were performed (e.g. missing flux, low S/N, resolved-out flux).  Such anomalies lead to asymmetrical or unusual features that are not related to intrinsic properties of the sources and are therefore still consistent with a spherically symmetric CSE.  As mentioned in Sect. 4.1, the S/N of the data of the weakest sources is relatively low (S/N$<$7). Because of the high level of noise, the emission distribution of weak sources cannot be properly fitted by a Gaussian, and the line profiles are of anomalous shapes. The lines from both CO transitions from the S-stars UY$\,\,$Cen and T$\,\,$Sgr suffer from relatively high noise levels. The carbon stars V1302$\,\,$Cen and V996$\,\,$Cen also present noisy data in both CO transitions. 

A consistent offset between the centre position of the fitted visibility in both CO(2--1) and CO(3--2) lines indicates that the coordinates of the source that were used to perform the observations, taken from the SIMBAD database, carry large uncertainties. This does not affect the brightness distribution. Such shifts in coordinates are observed in the S-stars NSV$\,\,$24833, and Z$\,\,$Ant. This behaviour is also observed in the C-star X$\,$TrA. 

Some of the sources are barely resolved and exhibit nearly flat size distributions (see Fig. \ref{fig:UV_fit_S1}--\ref{fig:UV_fit_MC}), that is, the sizes appear independent of the radial velocity in the
channel maps,  in contrast with the expected distribution following $\sqrt{1-((\upsilon-\upsilon_\textmd{sys})/\upsilon_\infty)^2}$, where $\upsilon_\textmd{sys}$ is the systemic velocity, and $\upsilon_\infty$ the terminal outflow velocity. The S-type star Z$\,\,$Ant and the C-type stars V996$\,\,$Cen and V1302$\,\,$Cen are barely resolved in CO(3--2). Observations at higher angular resolution are needed to better constrain the shapes of those sources.

As discussed in Sect. 4.2, some of the S-stars in our sample suffer from resolved-out flux. As the ACA fails to recover flux on large-scale structures, the intensity distribution of the largest sources cannot be well-fitted by a simple Gaussian. This affects the CO(3--2) line to a greater extent because of the smaller beam size. Missing flux is either indicated by large errors or by fluctuating spikes in the major and minor axes of the fitted Gaussian. These are observed in both CO transitions in DY Gem and ST$\,\,$Sgr, and in the $3\textrm{--}2$ line in FU$\,\,$Mon, GI$\,\,$Lup, IRC-10401, NSV$\,\,$24833, RT$\,\,$Sco, and V821$\,\,$Her. Zero-spacing observations are needed to recover the missing flux.

Self-absorption on the blue side of the line also alters the shape of the line profile. Self-absorption is indicated by a lower emerging flux on the blueshifted part of the line. This also affects the estimated size of the source which peaks in the blueshifted velocity channels as they probe an emitting volume that is larger 
overall. This effect is observed in the S-star RT$\,$Sco. Both CO lines of IRC-10401 are affected by confusion with interstellar CO emission that is not resolved out by the interferometer.
\subsubsection{Features indicative of anisotropy}
Some other interesting features are clearly indicative of circumstellar anisotropic structures. Such features are likely to indicate intrinsic deviation from spherical symmetry. This applies to triangular profiles; these are not yet well understood but are known to be common in high-MLR sources. The S-star FU$\,$Mon, with a previously estimated MLR of $2.7 \times 10^{-7}$ M$_\odot$ yr$^{-1}$, exhibits a triangular line profile. This is also the case of the C-type stars AFGL$\,$3068 ($1\times10^{-5}$ M$_\odot$ yr$^{-1}$) and U Men ($2\times 10^{-7}$ M$_\odot$ yr$^{-1}$). 

A number of the S-stars in our sample possess a profile with a wide base, which could be representative of fast outflows, and multiple peaks, which can sometimes be linked to the presence of an expanding torus \citep[e.g. $\pi$ Gru$^1$,][]{Lam2017}. In our sample, the sources presenting a wide base combined with some peculiar peak structure ---sometimes with a velocity symmetry--- are Z Ant, FU$\,\,$Mon, DY$\,\,$Gem, RZ$\,\,$Sgr, ST$\,\,$Sgr, Z$\,\,$Ant, and $\theta \,$Aps in both CO(2--1) and CO(3--2) lines. 

Rotation or plane-parallel expansion can possibly explain the asymmetric nature of the line profiles of some of our sources. This can be indicated by a position gradient with respect to the phase centre when consistently observed in both lines \citep[e.g.][]{sofia2018,Wouter2018,deathstar}. Such a central position gradient is observed in the S-stars DY Gem in RA, FU$\,\,$Mon in both RA and Dec, RZ$\,\,$Sgr in RA, T$\,\,$Cet and T$\,\,$Sgr in RA and Dec, and in the M-star R$\,\,$Dor in RA and Dec. 

Thanks to the high sensitivity offered by the ACA, unusual extended spiral structures are observed in the images of RZ$\,\,$Sgr and FU$\,\,$Mon, observable in both the 2--1 and 3--2 lines (Fig. \ref{fig:RZ_Sgr_channel21}--\ref{fig:FU_Mon_channel32}). For several sources (e.g. T$\,\,$Cet and UY$\,\,$Cen), the emission shows significant velocity asymmetry that is inconsistent with self-absorption but traces the dynamics of the CO gas (see Fig. \ref{fig:T_Cet_channel21}--\ref{fig:T_Cet_channel32} for the channels maps of T Cet). The asymmetries listed above are believed to be intrinsic to the sources and therefore cannot be consistent with a spherically symmetric CSE.

We indicate in the last columns of Table 2 if the observed deviations from spherical symmetry are due to external (observational setups) or to internal processes for all the sources in our sample.  Better observations, that is, either at higher resolution or sensitivity, are necessary for a more accurate representation of the shape of the sources that exhibit extrinsic asymmetrical features in order to draw pertinent conclusions as to whether they deviate from spherical symmetry or not. For the sources that present signs of intrinsic asymmetrical features, that is, not consistent with a smooth, symmetric outflow, further observations and modelling are needed to better understand the physical processes behind them.
 
\subsection{S-type CSEs larger than 3$\,$000 AU}
In this section, we investigate the possible reasons for the large measured sizes of the five largest S-type CSEs. For that, we look at how the large- and small-scale asymmetries of the emissions described in Sect. 4.3 could have affected the Gaussian fitting, and  could  therefore affect the derived sizes. We also discuss the possible influence of the adopted distances.
\subsubsection*{DY$\,$Gem}
DY$\,\,$Gem is an SRa S-type AGB star with a MLR of $7 \times 10^{-7}$ M$_\odot$ yr$^{-1}$. We estimate its CO FWHM size to be about 3$\,$100 AU.  The fitted Gaussian as a function of channel for this star presents fluctuating spikes in a few channels for both CO  lines, suggesting that the emission is resolved out in those channels. In particular, we observe an upward spike at the central velocity channel from which the CO(2--1) size is estimated. However, this is not likely to explain the obtained large size of the star, as the upward spike would only indicate that flux is missing in the channels surrounding the central velocity channel. Furthermore,  it is clear from the results of the visibility fitting that DY$\,$Gem  presents a central position gradient in RA, which could be caused by rotation or by the presence of an expanding torus. In addition, its line profile exhibits a wide base and two peaks in both the CO(2--1) and (3--2) lines. Its aspherical structure is also confirmed by its minor-to-major axis ratio of $\sim 0.51$. We adopt a distance of 680 pc for DY$\,$Gem \citep{Sofia2006,Sofia2009} to estimate its CO size. Its distance measured by Gaia eDR3 is 906 pc \citep{GaiaDR3}. Using this more recent distance gives an even larger linear size. 

The apparent anisotropic features and effect of observational limitations observed in the results of the visibility fitting of DY$\,$Gem undoubtedly affect the measured size. Whether the observed abnormalities lead to an overestimation or underestimation of the size of the CSE is not clear. The derived size is therefore not likely to represent the true size of the CO(2--1)-emitting region.
\subsubsection*{AM$\,$Cen and TT$\,$Cen}
AM$\,\,$Cen is an SRb star with a MLR of $1.5 \times 10^{-7}$ M$_\odot$ yr$^{-1}$. TT$\,\,$Cen is a Mira star losing mass at a rate of $2.5 \times 10^{-6}$ M$_\odot$ yr$^{-1}$. The measured CO FWHM sizes are 4$\,$200 and 5$\,$500 AU for AM$\,\,$Cen and TT$\,\,$Cen, respectively.  AM$\,$Cen presents an extended wing on the blue side of both the CO(2--1) and CO(3--2) lines.  

The CO line emissions from both stars are properly fitted by a Gaussian distribution, with no sign of intrinsic large-scale asymmetry. The Gaia eDR3 distances for both stars from the \citet{GaiaDR3} catalogue are  larger than the distances used to measure the CO sizes (see Table 1). Even larger linear CO sizes are derived for AM$\,\,$Cen and TT$\,\,$Cen when using the Gaia distances. 

The goodness of the fits and the absence of strong anisotropic features are both indicative of the reliability of the measurements of the CO(2--1)-emitting region of both stars. Assuming that the Gaia distances validate that the distances adopted in our calculations are lower limits, we can conclude that the large measured sizes of both stars are likely to be lower limits, which would imply that AM Cen and TT Cen have intrinsically large CSEs. 
\subsubsection*{FU$\,$Mon and RZ$\,$Sgr}
The largest sources in our sample of S-stars are FU$\,$Mon and RZ$\,$Sgr. FU Mon is a semi-regular star with a MLR of $2.7 \times 10^{-7}$ M$_\odot$ yr$^{-1}$. Its measured CO size is 6$\,$500 AU. RZ$\,$Sgr is an SRb star. It has the highest MLR, namely $3 \times 10^{-6}$ M$_\odot\,$yr$^{-1}$, and the largest measured CO size of 6$\,$600 AU in our S-stars sample. FU$\,$Mon exhibits a triangular-shaped profile with a wide base in both CO(2--1) and CO(3--2). RZ$\,$Sgr presents a two-horned line profile in both CO transitions. The results from the visibility fittings of the two CO lines show that both stars present a change in their centre position, along both RA and Dec for FU$\,$Mon, and along RA for RZ$\,$Sgr. The shape of the distribution of their major axis is similar. Their CO(2--1) line emission distributions are roughly fitted by a Gaussian, with signs of missing flux across the central velocity channels. The emission from both CO transitions is affected by resolved-out flux, but the drop in flux in the CO(3--2) line is particularly significant, resulting in a major axis distribution that has an approximately two-peaked shape. In the image plane, both stars clearly exhibit symmetrical extended structures in both the CO(2--1) and CO(3--2) lines (Figs \ref{fig:RZ_Sgr_channel21}--\ref{fig:FU_Mon_channel32}) that could indicate the presence of a torus, jets, or other physical phenomena. Further high-resolution observations are needed to understand the nature and origin of those structures. The distance to FU$\,\,$Mon used in this study  is of the order of its Gaia eDR3 distance. The Gaia eDR3 distance of RZ Sgr is smaller than the distance used in our calculations by a factor of 1.7, which, if correct, would imply that the measured size of RZ$\,\,$Sgr is overestimated by the same factor.
The evident aspherical structures present in RZ$\,\,$Sgr and FU$\,\,$Mon lead to uncertainties in the measured sizes, making them relatively unreliable. How these structures affect the size determination is not yet fully understood.

 \subsection{Detection of emission from molecules other than $^{12}$CO}
The frequencies covered by the observed bands permit the detection of a handful of molecules in addition to the $^{12}$CO lines needed to constrain the CSE sizes. The peak fluxes of the detected lines measured within a 10$^{\prime \prime}$ aperture are listed in Table \ref{tab:other_molecules} for the S-type sources. The main molecular lines detected in most of the sources are $^{13}$CO, CS, and oxygen-bearing molecules such as SiO and $^{29}$SiO. 

As S-stars are believed to be objects transitioning from M- to C-type, their C/O ratio can be slightly oxygen-rich or carbon-rich. According to chemical equilibrium calculations by \citet{Agudez2020} on high-MLR stars ($\sim 10^{-5}$ M$_\odot$ yr$^{-1}$) with a C/O of around 1, the atmosphere of S-type stars is relatively similar to that of C-stars. This is because C-bearing molecules are more efficient in competing with CO to get the carbon than O-bearing molecules for the oxygen, with the exception of SiO. Therefore, carbon-bearing molecules may be of moderate abundance even in slightly oxygen-rich environments. Furthermore, recent observations of the S-star W Aql (C/O $\simeq 0.98)$ by \citet{Elvire2020} show that the atmosphere of the S-star is more similar to the atmosphere of a C-type star than of an M-type star.  \citet{Elvire2020} reported the first detections of molecules that have only been detected in C-type stars including SiC$_2$, HC$_3$N, SiN, and C$_2$H in an S-type AGB star. We do not detect such molecules typical of C-rich environments in our S-stars at the given spectral resolution and aperture. The detection of SiO is expected in S-stars as it efficiently competes with CO to lock oxygen and is the most abundant O-bearing molecule for all chemical types \citep[e.g.][]{Sofia2009}. 

The lines in Table \ref{tab:other_molecules} are those for which we are confident of detection at the given spectral resolution and aperture. Further optimisation of the data analysis could lead to further detections, which is beyond the scope of this paper.
 \section{Discussion and summary}
The new ACA observations presented in this paper show that, for a similar density, the sizes of the CO CSEs of low-MLR S-type stars fall between the sizes of C-CSEs, which are larger, and M-CSEs which are smaller. This is likely related to the differences in their respective CO abundances through photodissociation. On the other hand, our results show that the CSEs of about 50\%\ of the high-MLR S-stars in our sample are larger than those of both C- and M-stars for the same density. The measured sizes of the low-density S-CSEs are smaller than expected based on the photodissociation results of \citet{Saberi2019}, whereas the high-density CSEs of S-stars appear to be much larger than the predicted CO(2--1) sizes. We performed a more detailed analysis of the properties of the five largest S-stars to understand whether their emission distributions have been properly fitted by a Gaussian. We find that three out of the five large sources display noticeable signs of intrinsic aspherical features and missing flux that affected the visibility fitting. Although their large derived sizes might be explained by, and not contradictory to, the presence of these large- and small-scale asymmetries, there is no further evidence to support that they lead to an overestimation of the measured sizes. In addition, two of the large sources are not as heavily impacted by asymmetries, and their emission distributions are properly fitted by a Gaussian. Further investigation is required in order to understand the physical and/or chemical reason behind the large derived sizes. The reliability of the Gaussian fit estimates decreases with increasing MLR which will affect the larger sources in particular \citep{deathstar}. This can be due to excitation and optical depth effects and/or to an intrinsic non-spherically smooth CSE. The measured sizes are also affected by the large uncertainties in the distances. Using the more recent Gaia distances listed in Table \ref{tab:sample} would lead to even larger CO CSEs for most of the sources (see \citealt{deathstar}). However, we use the previous distances in Fig. \ref{fig:size-density} for consistency, as the densities on the x-axis are derived from MLRs that are based on these older distances.  Replacing the values of the distances would not only change the derived sizes, but would also require a similar change in the MLR, and would therefore  not significantly affect how the derived sizes are related to the photodissociation diameters.  

As for the C- and M-stars in \citet{deathstar}, the CSE size distribution of the low-MLR stars presented in this paper shows no density dependence, irrespective of their variability types. A correlation between size and density is nevertheless observed  for high-MLR stars. The density dependence of the size is steeper for the S-stars compared to the other chemical types at high MLR.  

The difference between observation and theory could be due to a systematic overestimation of the CO abundance, or differences in the adopted parameters such as the temperature profile, the UV environment, the dust properties, or the density profiles. A number of the observations presented in this paper are subject to missing or resolved-out flux, or high noise levels that can significantly alter the Gaussian fits. Therefore, a Gaussian distribution is not always a good fit to the CO line emission distribution of AGB stars due to intrinsic properties of the sources or to the quality of observations, and the sizes measured using this method should be considered as a first approximation. The results presented here combined with single-dish data will serve as constraints to detailed radiative transfer modelling that will be presented in future publications.  

Our results show that the CO CSEs of most of the southern S-stars in the DEATHSTAR sample are consistent with a spherically symmetric and smooth outflow. This conclusion comes from our investigation of the possible signs of deviation from spherical symmetry based on line profiles, minor-to-major axis ratio from visibility fitting, and by looking at the deconvolved images. For some of the sources, clear and prominent asymmetric features are observed that are indicative of a more complex structure. Follow-up studies are needed to understand the effects of such deviations on the estimated MLRs. 

We report the detection of several molecules other than $^{12}$CO in our sample of S-type stars, namely $^{13}$CO, CS, SiO, and $^{29}$SiO. No further analysis of the emission from molecules other than $^{12}$CO  has been attempted. This will be done in the future within the scope of the DEATHSTAR project.

\begin{table*}[htbp]
\centering
  \caption{Sources by spectral type in ascending MLR order.}
    \resizebox{13cm}{!}{
    \begin{tabular}{l|lllll}
    \hline \hline
    Source & \multicolumn{1}{p{6.335em}}{Variability \newline{}type} & \multicolumn{1}{p{5em}}{$\dot{M}$ \newline{}[M$_\odot$ yr$^{-1}$]} & \multicolumn{1}{p{4em}}{$\upsilon_\infty$ \newline{}[km s$^{-1}$]} & \multicolumn{1}{p{3.165em}}{$D$\newline{}[pc]} & \multicolumn{1}{p{4em}}{$D_\textmd{gaia}^{5}$ \newline{}[pc]} \\
    \hline \hline
    \multicolumn{6}{l}{\textit{S-type semi-regulars and irregular stars$^1$:}} \\
    T Cet & SRc   & 4$\, \times \,10^{-8}$ & 5.5   & 240   & 252 \\
    Z Ant & SRb   & 9$\, \times \,10^{-8}$ & 6.0     & 470   & 936 \\
    UY Cen & SRb   & 1.30$\, \times \,10^{-7}$ & 12.0    & 590   & 664 \\
    AM Cen & SRb   & 1.50$\, \times \,10^{-7}$ & 3.2   & 750   & 907 \\
    ST Sco & SRa   & 1.50$\, \times \,10^{-7}$ & 5.5   & 380   & 623 \\
    FU Mon & SR    & 2.70$\, \times \,10^{-7}$ & 2.8   & 780   & 789 \\
    NSV 24833 & U & 3$\, \times \,10^{-7}$ & 8.7   & 610   & 1106 \\
    DY Gem & SRa   & 7$\, \times \,10^{-7}$ & 8.0     & 680   & 906 \\
    RZ Sgr & SRb   & 3$\, \times \,10^{-6}$ & 9.0    & 730   & 424 \\
    \multicolumn{6}{l}{\textit{S-type Mira stars$^1$:}} \\
    TT Cen & M     & 2.50$\, \times \,10^{-6}$ & 20.0    & 880   & 1348 \\
    GI Lup & M     & 5.50$\, \times \,10^{-7}$ & 10.0    & 690   & 1085 \\
    RT Sco & M     & 4.50$\, \times \,10^{-7}$ & 11.0    & 270   & 756 \\
    IRC-10401 & M    & 3.50$\, \times \,10^{-7}$ & 17.0    & 430   & 2062 \\
    ST Sgr & M     & 2$\, \times \,10^{-7}$ & 6.0     & 540   & 542 \\
    T Sgr & M     & 1.40$\, \times \,10^{-7}$ & 7.5   & 590   & 3922 \\
    W Aql & M     & 2.20$\, \times \,10^{-6}$ & 17.2  & 230   & 375 \\
    \multicolumn{6}{l}{\textit{M-type semi-regulars and irregular stars$^{2\, ,3 \, , 4}$:}} \\
    $\theta$ Aps & SRb   & 4$\, \times \,10^{-8}$ & 4.5   & 110   & 119 \\
    R Dor$^*$ & SRb   & 1.30$\, \times \,10^{-7}$ & 5.5   & 45    & - \\
    U Men & SRa   & 2$\, \times \,10^{-7}$ & 7.2   & 320   &325\\
    \multicolumn{6}{l}{\textit{M-type Mira stars$^{2\, ,3 \, , 4}$:}} \\
    IRC-30398$^*$ & M$\,$:    & 1.10$\, \times \,10^{-5}$ & 16.0    & 600   & 926 \\
    \multicolumn{6}{l}{\textit{C-type semi-regular and irregular stars$^{2\, ,3 \, , 4}$:}} \\
    V1302 Cen & SRb   & 1$\, \times \,10^{-7}$ & 6.5   & 530   & 911 \\
    V996 Cen & SR    & 1$\, \times \,10^{-7}$ & 11.0    & 390   & 576 \\
   X TrA & Lb    & 1.30$\, \times \,10^{-7}$ & 6.5   & 360   & 350 \\
    AQ Sgr & SRb   & 2.50$\, \times \,10^{-7}$ & 10.0    & 200   & -\\
    \multicolumn{6}{l}{\textit{C-type Mira stars$^{2\, ,3 \, , 4}$:}} \\
    V821 Her & M     & 3$\, \times \,10^{-6}$ & 13.0    & 600   & 752 \\
    AI Vol$^*$ & M     & 4.90$\, \times \,10^{-6}$ & 12.0    & 710   & 624 \\
    AFGL 3068$^*$ & M     & 1$\, \times \,10^{-5}$ & 13.5  & 980   &  -\\
    \bottomrule
    \multicolumn{4}{l}{\footnotesize $\dot{M}$ is the MLR, $\upsilon_\infty$ the terminal expansion velocity, $D$ the  distance } \\
    \multicolumn{4}{l}{\footnotesize * Observed with TP } \\
    \multicolumn{4}{l}{\footnotesize $^{1}$\cite{Sofia2009}  } \\
    \multicolumn{4}{l}{\footnotesize $^{2}$ \citet{schoier2001}} \\
    \multicolumn{4}{l}{\footnotesize $^{3}$\cite{Gonzalez2003}} \\
    \multicolumn{4}{l}{\footnotesize $^{4}$\cite{Danilovich2015}} \\
    \multicolumn{4}{l}{\footnotesize $^{5}$\cite{GaiaDR3, Bailer-Jones2021}.} \\
    \end{tabular}}%
  \label{tab:sample}%
\end{table*}%

\begin{table*}[htbp]
\centering
  \caption{Results from the Gaussian fitting\tnote{2}.}
    \begin{adjustbox}{width=15cm}
    \begin{tabular}{l|c|cc|cc|cc}
    \hline \hline
    & \multicolumn{1}{c|}{\multirow{2}{*}{$\frac{2R_p^{**}}{D}$}} & \multicolumn{2}{c}{CO (2-1)} & \multicolumn{2}{c|}{CO (3-2)} &  \multicolumn{2}{c}{Asymmetry$^{***}$}\\
    \cmidrule{3-8}     Source     &   [$^{\prime\prime}$]    & \multicolumn{1}{p{5.6em}}{Major axis\newline{}[$^{\prime\prime}$]} & \multicolumn{1}{p{5.2em}}{Axes ratio\newline{}} & \multicolumn{1}{p{5.3em}}{Major axis\newline{}[$^{\prime\prime}$]} & \multicolumn{1}{p{5.2em}|}{Axes ratio\newline{}} & \multicolumn{1}{p{3em}}{Extrinsic\newline{}} & \multicolumn{1}{p{3em}}{Intrinsic\newline{}}
    \\
\hline \hline
    \multicolumn{6}{l}{\textit{S-type semi-regulars and irregular stars:}} \\
    T Cet & 10.8  & \multicolumn{1}{c}{4.58$\pm$0.34 } & \multicolumn{1}{c}{0.79$\pm$0.06} & 3.42$\pm$0.07 & 0.83$\pm$0.02 & \multicolumn{1}{c|}{-} &  \multicolumn{1}{c|}{$g$, $h$}\\
    Z Ant & 6.2   & \multicolumn{1}{c}{3.24$\pm$0.17} & 0.84$\pm$0.06 & 1.83$\pm$0.11 & 0.91$\pm$0.25 & \multicolumn{1}{c|}{$b$} &  \multicolumn{1}{c|}{$f$} \\
    UY Cen & 5.0   & \multicolumn{1}{c}{1.49$\pm$0.71} & 0.52$\pm$0.23 & 1.39$\pm$0.55 & 0.98$\pm$0.42 & \multicolumn{1}{c|}{$c$}& \multicolumn{1}{c|}{$g$} \\
    AM Cen & 6.8   & \multicolumn{1}{c}{5.59$\pm$0.27} & 0.90$\pm$0.06 & 2.16$\pm$0.09 & 0.86$\pm$0.07 & \multicolumn{1}{c|}{-}& \multicolumn{1}{c|}{-}\\
    ST Sco & 10.6  & \multicolumn{1}{c}{4.32$\pm$0.10} & 1.00$\pm$0.02 & 2.99$\pm$0.08 & 0.92$\pm$0.04 & \multicolumn{1}{c|}{-}& \multicolumn{1}{c|}{-}\\
    FU Mon & 10.2  & \multicolumn{1}{c}{8.29$\pm$0.17} & 0.84$\pm$0.03 & 2.36$\pm$0.19 & 0.80$\pm$0.07 & \multicolumn{1}{c|}{-} &  \multicolumn{1}{c|}{$e$, $g$, $h$}\\
    NSV 24833 & 8.2   & \multicolumn{1}{c}{3.76$\pm$0.21} & 0.96$\pm$0.06 & 2.35$\pm$0.09 & 1.00$\pm$0.05 & \multicolumn{1}{c|}{-} & \multicolumn{1}{c|}{-}\\
    DY Gem & 12.7  & \multicolumn{1}{c}{4.50$\pm$0.15} & 0.51$\pm$0.04 & 2.49$\pm$0.15 & 0.78$\pm$0.06 & \multicolumn{1}{c|}{$a$}& \multicolumn{1}{c|}{$f$, $g$, $h$}\\
    RZ Sgr & 27.6  & \multicolumn{1}{c}{9.00$\pm$0.09} & 0.99$\pm$0.01 & 3.36$\pm$0.04 & 0.94$\pm$0.01 & \multicolumn{1}{c|}{-} & \multicolumn{1}{c|}{$f$, $g$} \\
    \multicolumn{6}{l}{\textit{S-type Mira stars:}} \\
    TT Cen & 14.2  & \multicolumn{1}{c}{6.26$\pm$0.31} & 1.00$\pm$0.07 & 5.39$\pm$0.10 & 0.90$\pm$0.02 &\multicolumn{1}{c|}{-} & \multicolumn{1}{c|}{-}\\
    GI Lup & 9.8   & \multicolumn{1}{c}{3.81$\pm$0.14} & 0.95$\pm$0.06 & 2.47$\pm$0.19 & 0.91$\pm$0.08 & \multicolumn{1}{c|}{-} & \multicolumn{1}{c|}{-}\\
    RT Sco & 21.6  & \multicolumn{1}{c}{6.99$\pm$0.07} & 1.00$\pm$0.01 & 4.09$\pm$0.07 & 1.00$\pm$0.01 & \multicolumn{1}{c|}{-}& \multicolumn{1}{c|}{-}\\
    IRC -10401 & 10.6  & \multicolumn{1}{c}{5.43$\pm$0.42} & 0.85$\pm$0.06 & 0.72: & 0.91$\pm$0.77 &\multicolumn{1}{c|}{-} &  \multicolumn{1}{c|}{-}\\
    ST Sgr & 8.5   & \multicolumn{1}{c}{1.52$\pm$0.17} & 0.98$\pm$0.05 & 1.27$\pm$0.09 & 0.89$\pm$0.03 &\multicolumn{1}{c|}{$a$, $c$, $d$} & \multicolumn{1}{c|}{$g$}\\
    T Sgr & 5.8   & \multicolumn{1}{c}{3.55$\pm$0.40} & 0.99$\pm$0.10 & 1.80$\pm$0.13 & 0.99$\pm$0.16 &\multicolumn{1}{c|}{$c$} & \multicolumn{1}{c|}{$g$}\\
    W Aql &   46.5    &    -   &   -    & 4.62$\pm$0.05 & 1.00$\pm$0.02 & \multicolumn{1}{c|}{-} &  \multicolumn{1}{c|}{$f$, $g$}\\
    \multicolumn{6}{l}{\textit{M-type semi-regulars and irregular stars:}} \\
    $\theta$ Aps & \multicolumn{1}{r|}{10.1} & 4.90$\pm$0.06 & 0.97$\pm$0.01 & 3.52$\pm$0.02 & 0.98$\pm$0.01 &\multicolumn{1}{c|}{-} & \multicolumn{1}{c|}{$f$}\\
    R Dor$^*$ & \multicolumn{1}{r|}{42.7} & \multicolumn{1}{c}{14.69$\pm$0.27}&0.92$\pm$0.03& 8.44: & 1.00:  &\multicolumn{1}{c|}{-} & \multicolumn{1}{c|}{$f$, $g$}\\
    U Men & \multicolumn{1}{r|}{7.0} & 7.15$\pm$0.20 & 0.88$\pm$0.03 & 3.56$\pm$0.12 & 0.94$\pm$0.03  &\multicolumn{1}{c|}{-} &  \multicolumn{1}{c|}{$f$, $g$}\\
    \multicolumn{6}{l}{\textit{M-type Mira stars:}} \\
    IRC-30398$^*$ & 27.5  & \multicolumn{1}{c}{12.24$\pm$0.41}&0.82$\pm$0.06& 4.85$\pm$0.05 & 0.88$\pm$0.02 &\multicolumn{1}{c|}{-} & \multicolumn{1}{c|}{-}\\
    \multicolumn{6}{l}{\textit{C-type semi-regular and irregular stars:}} \\
    V1302 Cen & 6.3   & \multicolumn{1}{c}{1.94$\pm$0.28} & 1.00$\pm$0.11 & 1.08$\pm$0.31 & 0.98$\pm$0.07 & \multicolumn{1}{c|}{$b$, $c$}& \multicolumn{1}{c|}{-} \\
    V996 Cen & 7.4   & \multicolumn{1}{c}{4.58$\pm$0.27} & 0.85$\pm$0.06 & 2.38$\pm$0.34 & 0.97$\pm$0.14 & \multicolumn{1}{c|}{$b$, $c$}&\multicolumn{1}{c|}{-}\\
    X TrA & 10.7  & \multicolumn{1}{c}{6.19$\pm$0.06} & 0.99$\pm$0.01 & 0.72$\pm$0.34 & 0.88$\pm$0.42 &\multicolumn{1}{c|}{-} & \multicolumn{1}{c|}{-}\\
    AQ Sgr & 11.5  &  -     &    -   & 0.93$\pm$0.36 & 0.90$\pm$0.04 &\multicolumn{1}{c|}{$a$} &\multicolumn{1}{c|}{-}\\
    \multicolumn{6}{l}{\textit{C-type Mira stars:}} \\
    V821 Her & 31.5  & \multicolumn{1}{c}{6.53$\pm$0.04} & 0.98$\pm$0.01 & 1.36$\pm$0.34 & 0.96$\pm$0.03 & \multicolumn{1}{c|}{$a$}&\multicolumn{1}{c|}{-}\\
    AI Vol$^*$ & 37.6  & \multicolumn{1}{c}{14.56$\pm$0.60}&0.94$\pm$0.06& 3.85$\pm$0.03 & 0.95$\pm$0.01 &\multicolumn{1}{c|}{-} &  \multicolumn{1}{c|}{-}\\
    AFGL 3068$^*$ & 40.3  & \multicolumn{1}{c}{17.14$\pm$0.73}&0.86$\pm$0.06& 4.12$\pm$0.04 & 0.94$\pm$0.01 &\multicolumn{1}{c|}{-} & \multicolumn{1}{c|}{$e$}\\
    \addlinespace
    \bottomrule
    \multicolumn{4}{l}{\footnotesize * Observed with TP} \\
    \multicolumn{4}{l}{\footnotesize **$R_p$ is the photodissociation radius.} \\
    \multicolumn{4}{l}{\footnotesize ***$a$: missing flux in both transitions} \\
    \multicolumn{4}{l}{\footnotesize $b$: unresolved} \\
    \multicolumn{4}{l}{\footnotesize $c$: noisy} \\
    \multicolumn{4}{l}{\footnotesize $d$: self absorption} \\
     \multicolumn{4}{l}{\footnotesize $e$: triangular} \\
     \multicolumn{4}{l}{\footnotesize $f$: wide base and multiple peaks} \\
     \multicolumn{4}{l}{\footnotesize $g$: position gradient} \\
     \multicolumn{4}{l}{\footnotesize $h$: low value or irregular behaviour of the major/minor axes ratio.} \\
    \end{tabular}
    \label{tab:Visibility}%
        \end{adjustbox}
\end{table*}%
 \begin{acknowledgements}
 This paper makes use of the following ALMA data: ADS/JAO.ALMA\#2018.1.01434.S; ADS/JAO.ALMA\#2017.1.00595.S and ADS/JAO.ALMA\#2012.1.00524.S.  ALMA is a partnership of ESO (representing its member states), NSF (USA) and NINS (Japan), together with NRC (Canada), MOST and ASIAA (Taiwan), and KASI (Republic of Korea), in cooperation with the Republic of Chile. The Joint ALMA Observatory is operated by ESO, AUI/NRAO and NAOJ. MA acknowledges support from the Nordic ALMA Regional Centre (ARC) node based at Onsala Space Observatory. The Nordic ARC node is funded through Swedish Research Council grant No 2017-00648. This project has received funding from the European Research Council (ERC) under the European Union’s Horizon 2020 research and innovation programme under grant agreements No. 883867 [EXWINGS] and 730562 [RadioNet].
\end{acknowledgements}
\bibliography{40952_Deathstar}

\begin{thebibliography}{41}
\expandafter\ifx\csname natexlab\endcsname\relax\def\natexlab#1{#1}\fi

\bibitem[{{Ag{\'u}ndez} {et~al.}(2020){Ag{\'u}ndez}, {Mart{\'\i}nez}, {de
  Andres}, {Cernicharo}, \& {Mart{\'\i}n-Gago}}]{Agudez2020}
{Ag{\'u}ndez}, M., {Mart{\'\i}nez}, J.~I., {de Andres}, P.~L., {Cernicharo},
  J., \& {Mart{\'\i}n-Gago}, J.~A. 2020, \aap, 637, A59

\bibitem[{{Bailer-Jones} {et~al.}(2021){Bailer-Jones}, {Rybizki}, {Fouesneau},
  {Demleitner}, \& {Andrae}}]{Bailer-Jones2021}
{Bailer-Jones}, C.~A.~L., {Rybizki}, J., {Fouesneau}, M., {Demleitner}, M., \&
  {Andrae}, R. 2021, \aj, 161, 147

\bibitem[{{Bladh} {et~al.}(2019){Bladh}, {Liljegren}, {H{\"o}fner}, {Aringer},
  \& {Marigo}}]{Bladh2019}
{Bladh}, S., {Liljegren}, S., {H{\"o}fner}, S., {Aringer}, B., \& {Marigo}, P.
  2019, \aap, 626, A100

\bibitem[{{Bl{\"o}cker}(1995)}]{bloecker1995}
{Bl{\"o}cker}, T. 1995, \aap, 297, 727

\bibitem[{{Cernicharo} {et~al.}(2015){Cernicharo}, {Marcelino}, {Ag{\'u}ndez},
  \& {Gu{\'e}lin}}]{Cernicharo2015}
{Cernicharo}, J., {Marcelino}, N., {Ag{\'u}ndez}, M., \& {Gu{\'e}lin}, M. 2015,
  \aap, 575, A91

\bibitem[{{Danilovich} {et~al.}(2015){Danilovich}, {Teyssier}, {Justtanont},
  {Olofsson}, {Cerrigone}, {Bujarrabal}, {Alcolea}, {Cernicharo},
  {Castro-Carrizo}, {Garc{\'\i}a-Lario}, \& {Marston}}]{Danilovich2015}
{Danilovich}, T., {Teyssier}, D., {Justtanont}, K., {et~al.} 2015, \aap, 581,
  A60

\bibitem[{{De Beck} {et~al.}(2010){De Beck}, {Decin}, {de Koter}, {Justtanont},
  {Verhoelst}, {Kemper}, \& {Menten}}]{Elvire2010}
{De Beck}, E., {Decin}, L., {de Koter}, A., {et~al.} 2010, \aap, 523, A18

\bibitem[{{De Beck} \& {Olofsson}(2020)}]{Elvire2020}
{De Beck}, E. \& {Olofsson}, H. 2020, \aap, 642, A20

\bibitem[{{Decin} {et~al.}(2019){Decin}, {Homan}, {Danilovich}, {de Koter},
  {Engels}, {Waters}, {Muller}, {Gielen}, {Garc{\'\i}a-Hern{\'a}ndez},
  {Stancliffe}, {Van de Sande}, {Molenberghs}, {Kerschbaum}, {Zijlstra}, \& {El
  Mellah}}]{Decin2019}
{Decin}, L., {Homan}, W., {Danilovich}, T., {et~al.} 2019, Nature Astronomy, 3,
  408

\bibitem[{{Decin} {et~al.}(2006){Decin}, {Hony}, {de Koter}, {Justtanont},
  {Tielens}, \& {Waters}}]{Decin2006}
{Decin}, L., {Hony}, S., {de Koter}, A., {et~al.} 2006, \aap, 456, 549

\bibitem[{{Doan} {et~al.}(2017){Doan}, {Ramstedt}, {Vlemmings}, {H{\"o}fner},
  {De Beck}, {Kerschbaum}, {Lindqvist}, {Maercker}, {Mohamed}, {Paladini}, \&
  {Wittkowski}}]{Lam2017}
{Doan}, L., {Ramstedt}, S., {Vlemmings}, W.~H.~T., {et~al.} 2017, \aap, 605,
  A28

\bibitem[{{Eriksson} {et~al.}(2014){Eriksson}, {Nowotny}, {H{\"o}fner},
  {Aringer}, \& {Wachter}}]{Eriksson2014}
{Eriksson}, K., {Nowotny}, W., {H{\"o}fner}, S., {Aringer}, B., \& {Wachter},
  A. 2014, \aap, 566, A95

\bibitem[{{Forestini} \& {Charbonnel}(1997)}]{Forestini1997}
{Forestini}, M. \& {Charbonnel}, C. 1997, \aaps, 123, 241

\bibitem[{{Gaia Collaboration} {et~al.}(2020){Gaia Collaboration}, {Smart},
  {Sarro}, {Rybizki}, {Reyl{\'e}}, {Robin}, {Hambly}, {Abbas}, {Barstow}, {de
  Bruijne}, {Bucciarelli}, {Carrasco}, {Cooper}, {Hodgkin}, {Masana},
  {Michalik}, {Sahlmann}, {Sozzetti}, {Brown}, {Vallenari}, {Prusti},
  {Babusiaux}, {Biermann}, {Creevey}, {Evans}, {Eyer}, {Hutton}, {Jansen},
  {Jordi}, {Klioner}, {Lammers}, {Lindegren}, {Luri}, {Mignard}, {Panem},
  {Pourbaix}, {Randich}, {Sartoretti}, {Soubiran}, {Walton}, {Arenou},
  {Bailer-Jones}, {Bastian}, {Cropper}, {Drimmel}, {Katz}, {Lattanzi}, {van
  Leeuwen}, {Bakker}, {Casta{\~n}eda}, {De Angeli}, {Ducourant}, {Fabricius},
  {Fouesneau}, {Fr{\'e}mat}, {Guerra}, {Guerrier}, {Guiraud}, {Jean-Antoine
  Piccolo}, {Messineo}, {Mowlavi}, {Nicolas}, {Nienartowicz}, {Pailler},
  {Panuzzo}, {Riclet}, {Roux}, {Seabroke}, {Sordo}, {Tanga}, {Th{\'e}venin},
  {Gracia-Abril}, {Portell}, {Teyssier}, {Altmann}, {Andrae}, {Bellas-Velidis},
  {Benson}, {Berthier}, {Blomme}, {Brugaletta}, {Burgess}, {Busso}, {Carry},
  {Cellino}, {Cheek}, {Clementini}, {Damerdji}, {Davidson}, {Delchambre},
  {Dell'Oro}, {Fern{\'a}ndez-Hern{\'a}ndez}, {Galluccio}, {Garc{\'\i}a-Lario},
  {Garcia-Reinaldos}, {Gonz{\'a}lez-N{\'u}{\~n}ez}, {Gosset}, {Haigron},
  {Halbwachs}, {Harrison}, {Hatzidimitriou}, {Heiter}, {Hern{\'a}ndez},
  {Hestroffer}, {Holl}, {Jan{\ss}en}, {Jevardat de Fombelle}, {Jordan},
  {Krone-Martins}, {Lanzafame}, {L{\"o}ffler}, {Lorca}, {Manteiga}, {Marchal},
  {Marrese}, {Moitinho}, {Mora}, {Muinonen}, {Osborne}, {Pancino}, {Pauwels},
  {Recio-Blanco}, {Richards}, {Riello}, {Rimoldini}, {Roegiers}, {Siopis},
  {Smith}, {Ulla}, {Utrilla}, {van Leeuwen}, {van Reeven}, {Abreu Aramburu},
  {Accart}, {Aerts}, {Aguado}, {Ajaj}, {Altavilla}, {{\'A}lvarez}, {{\'A}lvarez
  Cid-Fuentes}, {Alves}, {Anderson}, {Anglada Varela}, {Antoja}, {Audard},
  {Baines}, {Baker}, {Balaguer-N{\'u}{\~n}ez}, {Balbinot}, {Balog}, {Barache},
  {Barbato}, {Barros}, {Bartolom{\'e}}, {Bassilana}, {Bauchet},
  {Baudesson-Stella}, {Becciani}, {Bellazzini}, {Bernet}, {Bertone}, {Bianchi},
  {Blanco-Cuaresma}, {Boch}, {Bombrun}, {Bossini}, {Bouquillon}, {Bragaglia},
  {Bramante}, {Breedt}, {Bressan}, {Brouillet}, {Burlacu}, {Busonero},
  {Butkevich}, {Buzzi}, {Caffau}, {Cancelliere}, {C{\'a}novas},
  {Cantat-Gaudin}, {Carballo}, {Carlucci}, {Carnerero}, {Casamiquela},
  {Castellani}, {Castro-Ginard}, {Castro Sampol}, {Chaoul}, {Charlot},
  {Chemin}, {Chiavassa}, {Cioni}, {Comoretto}, {Cornez}, {Cowell}, {Crifo},
  {Crosta}, {Crowley}, {Dafonte}, {Dapergolas}, {David}, {David}, {de Laverny},
  {De Luise}, {De March}, {De Ridder}, {de Souza}, {de Teodoro}, {de Torres},
  {del Peloso}, {del Pozo}, {Delgado}, {Delgado}, {Delisle}, {Di Matteo},
  {Diakite}, {Diener}, {Distefano}, {Dolding}, {Eappachen}, {Edvardsson},
  {Enke}, {Esquej}, {Fabre}, {Fabrizio}, {Faigler}, {Fedorets}, {Fernique},
  {Fienga}, {Figueras}, {Fouron}, {Fragkoudi}, {Fraile}, {Franke}, {Gai},
  {Garabato}, {Garcia-Gutierrez}, {Garc{\'\i}a-Torres}, {Garofalo}, {Gavras},
  {Gerlach}, {Geyer}, {Giacobbe}, {Gilmore}, {Girona}, {Giuffrida}, {Gomel},
  {Gomez}, {Gonzalez-Santamaria}, {Gonz{\'a}lez-Vidal}, {Granvik},
  {Guti{\'e}rrez-S{\'a}nchez}, {Guy}, {Hauser}, {Haywood}, {Helmi}, {Hidalgo},
  {Hilger}, {H{\l}adczuk}, {Hobbs}, {Holland}, {Huckle}, {Jasniewicz},
  {Jonker}, {Juaristi Campillo}, {Julbe}, {Karbevska}, {Kervella}, {Khanna},
  {Kochoska}, {Kontizas}, {Kordopatis}, {Korn}, {Kostrzewa-Rutkowska},
  {Kruszy{\'n}ska}, {Lambert}, {Lanza}, {Lasne}, {Le Campion}, {Le Fustec},
  {Lebreton}, {Lebzelter}, {Leccia}, {Leclerc}, {Lecoeur-Taibi}, {Liao},
  {Licata}, {Lindstr{\o}m}, {Lister}, {Livanou}, {Lobel}, {Madrero Pardo},
  {Managau}, {Mann}, {Marchant}, {Marconi}, {Marcos Santos}, {Marinoni},
  {Marocco}, {Marshall}, {Polo}, {Mart{\'\i}n-Fleitas}, {Masip}, {Massari},
  {Mastrobuono-Battisti}, {Mazeh}, {McMillan}, {Messina}, {Millar}, {Mints},
  {Molina}, {Molinaro}, {Moln{\'a}r}, {Montegriffo}, {Mor}, {Morbidelli},
  {Morel}, {Morris}, {Mulone}, {Munoz}, {Muraveva}, {Murphy}, {Musella},
  {Noval}, {Ord{\'e}novic}, {Orr{\`u}}, {Osinde}, {Pagani}, {Pagano},
  {Palaversa}, {Palicio}, {Panahi}, {Pawlak}, {Pe{\~n}alosa Esteller},
  {Penttil{\"a}}, {Piersimoni}, {Pineau}, {Plachy}, {Plum}, {Poggio},
  {Poretti}, {Poujoulet}, {Pr{\v{s}}a}, {Pulone}, {Racero}, {Ragaini},
  {Rainer}, {Raiteri}, {Rambaux}, {Ramos}, {Ramos-Lerate}, {Re Fiorentin},
  {Regibo}, {Ripepi}, {Riva}, {Rixon}, {Robichon}, {Robin}, {Roelens},
  {Rohrbasser}, {Romero-G{\'o}mez}, {Rowell}, {Royer}, {Rybicki}, {Sadowski},
  {Sagrist{\`a} Sell{\'e}s}, {Salgado}, {Salguero}, {Samaras}, {Sanchez
  Gimenez}, {Sanna}, {Santove{\~n}a}, {Sarasso}, {Schultheis}, {Sciacca},
  {Segol}, {Segovia}, {S{\'e}gransan}, {Semeux}, {Shahaf}, {Siddiqui},
  {Siebert}, {Siltala}, {Slezak}, {Solano}, {Solitro}, {Souami}, {Souchay},
  {Spagna}, {Spoto}, {Steele}, {Steidelm{\"u}ller}, {Stephenson},
  {S{\"u}veges}, {Szabados}, {Szegedi-Elek}, {Taris}, {Tauran}, {Taylor},
  {Teixeira}, {Thuillot}, {Tonello}, {Torra}, {Torra}, {Turon}, {Unger},
  {Vaillant}, {van Dillen}, {Vanel}, {Vecchiato}, {Viala}, {Vicente},
  {Voutsinas}, {Weiler}, {Wevers}, {Wyrzykowski}, {Yoldas}, {Yvard}, {Zhao},
  {Zorec}, {Zucker}, {Zurbach}, \& {Zwitter}}]{GaiaDR3}
{Gaia Collaboration}, {Smart}, R.~L., {Sarro}, L.~M., {et~al.} 2020, arXiv
  e-prints, arXiv:2012.02061

\bibitem[{{Gonz{\'a}lez Delgado} {et~al.}(2003){Gonz{\'a}lez Delgado},
  {Olofsson}, {Kerschbaum}, {Sch{\"o}ier}, {Lindqvist}, \&
  {Groenewegen}}]{Gonzalez2003}
{Gonz{\'a}lez Delgado}, D., {Olofsson}, H., {Kerschbaum}, F., {et~al.} 2003,
  \aap, 411, 123

\bibitem[{{Habing}(1996)}]{Habing1996}
{Habing}, H.~J. 1996, \aapr, 7, 97

\bibitem[{{H{\"o}fner}(2015)}]{susanne2015}
{H{\"o}fner}, S. 2015, in Astronomical Society of the Pacific Conference
  Series, Vol. 497, Why Galaxies Care about AGB Stars III: A Closer Look in
  Space and Time, ed. F.~{Kerschbaum}, R.~F. {Wing}, \& J.~{Hron}, 333

\bibitem[{{H{\"o}fner} \& {Olofsson}(2018)}]{Hofner2018}
{H{\"o}fner}, S. \& {Olofsson}, H. 2018, \aapr, 26, 1

\bibitem[{{Jorissen} \& {Knapp}(1998)}]{jorissen1998}
{Jorissen}, A. \& {Knapp}, G.~R. 1998, \aaps, 129, 363

\bibitem[{{Mamon} {et~al.}(1988){Mamon}, {Glassgold}, \& {Huggins}}]{Mamon1988}
{Mamon}, G.~A., {Glassgold}, A.~E., \& {Huggins}, P.~J. 1988, \apj, 328, 797

\bibitem[{{Mart{\'\i}-Vidal} {et~al.}(2014){Mart{\'\i}-Vidal}, {Vlemmings},
  {Muller}, \& {Casey}}]{uvmultifit}
{Mart{\'\i}-Vidal}, I., {Vlemmings}, W.~H.~T., {Muller}, S., \& {Casey}, S.
  2014, \aap, 563, A136

\bibitem[{{Matsuura} {et~al.}(2009){Matsuura}, {Barlow}, {Zijlstra},
  {Whitelock}, {Cioni}, {Groenewegen}, {Volk}, {Kemper}, {Kodama}, {Lagadec},
  {Meixner}, {Sloan}, \& {Srinivasan}}]{Matsuura2009}
{Matsuura}, M., {Barlow}, M.~J., {Zijlstra}, A.~A., {et~al.} 2009, \mnras, 396,
  918

\bibitem[{{McDonald} {et~al.}(2015){McDonald}, {Zijlstra}, {Lagadec}, {Sloan},
  {Boyer}, {Matsuura}, {Smith}, {Smith}, {Yates}, {van Loon}, {Jones},
  {Ramstedt}, {Avison}, {Justtanont}, {Olofsson}, {Blommaert}, {Goldman}, \&
  {Groenewegen}}]{sunburn2015}
{McDonald}, I., {Zijlstra}, A.~A., {Lagadec}, E., {et~al.} 2015, \mnras, 453,
  4324

\bibitem[{{McMullin} {et~al.}(2007){McMullin}, {Waters}, {Schiebel}, {Young},
  \& {Golap}}]{casa}
{McMullin}, J.~P., {Waters}, B., {Schiebel}, D., {Young}, W., \& {Golap}, K.
  2007, in Astronomical Society of the Pacific Conference Series, Vol. 376,
  Astronomical Data Analysis Software and Systems XVI, ed. R.~A. {Shaw},
  F.~{Hill}, \& D.~J. {Bell}, 127

\bibitem[{Olofsson(1999)}]{olofsson_1999}
Olofsson, H. 1999, Symposium - International Astronomical Union, 191, 3–18

\bibitem[{{Olofsson}(2005)}]{Olofsson2005}
{Olofsson}, H. 2005, in ESA Special Publication, Vol. 577, ESA Special
  Publication, ed. A.~{Wilson}, 223--228

\bibitem[{{Olofsson} {et~al.}(2002){Olofsson}, {Gonz{\'a}lez Delgado},
  {Kerschbaum}, \& {Sch{\"o}ier}}]{Olofsson2002}
{Olofsson}, H., {Gonz{\'a}lez Delgado}, D., {Kerschbaum}, F., \& {Sch{\"o}ier},
  F.~L. 2002, \aap, 391, 1053

\bibitem[{{Ramstedt} {et~al.}(2018){Ramstedt}, {Mohamed}, {Olander},
  {Vlemmings}, {Khouri}, \& {Liljegren}}]{sofia2018}
{Ramstedt}, S., {Mohamed}, S., {Olander}, T., {et~al.} 2018, \aap, 616, A61

\bibitem[{{Ramstedt} {et~al.}(2017){Ramstedt}, {Mohamed}, {Vlemmings},
  {Danilovich}, {Brunner}, {De Beck}, {Humphreys}, {Lindqvist}, {Maercker},
  {Olofsson}, {Kerschbaum}, \& {Quintana-Lacaci}}]{Sofia2017}
{Ramstedt}, S., {Mohamed}, S., {Vlemmings}, W.~H.~T., {et~al.} 2017, \aap, 605,
  A126

\bibitem[{{Ramstedt} {et~al.}(2009){Ramstedt}, {Sch{\"o}ier}, \&
  {Olofsson}}]{Sofia2009}
{Ramstedt}, S., {Sch{\"o}ier}, F.~L., \& {Olofsson}, H. 2009, \aap, 499, 515

\bibitem[{{Ramstedt} {et~al.}(2006){Ramstedt}, {Sch{\"o}ier}, {Olofsson}, \&
  {Lundgren}}]{Sofia2006}
{Ramstedt}, S., {Sch{\"o}ier}, F.~L., {Olofsson}, H., \& {Lundgren}, A.~A.
  2006, \aap, 454, L103

\bibitem[{{Ramstedt} {et~al.}(2008){Ramstedt}, {Sch{\"o}ier}, {Olofsson}, \&
  {Lundgren}}]{Sofia2008}
{Ramstedt}, S., {Sch{\"o}ier}, F.~L., {Olofsson}, H., \& {Lundgren}, A.~A.
  2008, \aap, 487, 645

\bibitem[{{Ramstedt} {et~al.}(2020){Ramstedt}, {Vlemmings}, {Doan},
  {Danilovich}, {Lindqvist}, {Saberi}, {Olofsson}, {De Beck}, {Groenewegen},
  {H{\"o}fner}, {Kastner}, {Kerschbaum}, {Khouri}, {Maercker}, {Montez},
  {Quintana-Lacaci}, {Sahai}, {Tafoya}, \& {Zijlstra}}]{deathstar}
{Ramstedt}, S., {Vlemmings}, W.~H.~T., {Doan}, L., {et~al.} 2020, \aap, 640,
  A133

\bibitem[{{Saberi} {et~al.}(2019){Saberi}, {Vlemmings}, \& {De
  Beck}}]{Saberi2019}
{Saberi}, M., {Vlemmings}, W.~H.~T., \& {De Beck}, E. 2019, \aap, 625, A81

\bibitem[{{Samus'} {et~al.}(2017){Samus'}, {Kazarovets}, {Durlevich},
  {Kireeva}, \& {Pastukhova}}]{GCVS}
{Samus'}, N.~N., {Kazarovets}, E.~V., {Durlevich}, O.~V., {Kireeva}, N.~N., \&
  {Pastukhova}, E.~N. 2017, Astronomy Reports, 61, 80

\bibitem[{{Sch{\"o}ier} \& {Olofsson}(2001)}]{schoier2001}
{Sch{\"o}ier}, F.~L. \& {Olofsson}, H. 2001, \aap, 368, 969

\bibitem[{{Sch{\"o}ier} {et~al.}(2013){Sch{\"o}ier}, {Ramstedt}, {Olofsson},
  {Lindqvist}, {Bieging}, \& {Marvel}}]{schoier2013}
{Sch{\"o}ier}, F.~L., {Ramstedt}, S., {Olofsson}, H., {et~al.} 2013, \aap, 550,
  A78

\bibitem[{Stephenson(1984)}]{catalogueS}
Stephenson, C.~B. 1984, A general catalogue of galactic S stars, 2nd edn.,
  Publications of the Warner and Swasey Observatory ; v. 3, no. 1 (Cleveland,
  Ohio: Case Western Reserve University)

\bibitem[{{Van Eck} {et~al.}(2017){Van Eck}, {Neyskens}, {Jorissen}, {Plez},
  {Edvardsson}, {Eriksson}, {Gustafsson}, {J{\o}rgensen}, \&
  {Nordlund}}]{VanEck2017}
{Van Eck}, S., {Neyskens}, P., {Jorissen}, A., {et~al.} 2017, \aap, 601, A10

\bibitem[{{Vlemmings} {et~al.}(2018){Vlemmings}, {Khouri}, {De Beck},
  {Olofsson}, {Garc{\'\i}a-Segura}, {Villaver}, {Baudry}, {Humphreys},
  {Maercker}, \& {Ramstedt}}]{Wouter2018}
{Vlemmings}, W.~H.~T., {Khouri}, T., {De Beck}, E., {et~al.} 2018, \aap, 613,
  L4

\bibitem[{{Wachter} {et~al.}(2002){Wachter}, {Schr{\"o}der}, {Winters},
  {Arndt}, \& {Sedlmayr}}]{watcher2002}
{Wachter}, A., {Schr{\"o}der}, K.~P., {Winters}, J.~M., {Arndt}, T.~U., \&
  {Sedlmayr}, E. 2002, \aap, 384, 452

\end{thebibliography}
\begin{appendix}
\section{Imaging results.}
\begin{table*}[!htbp]
  \caption{Imaging results.}
\setlength{\tabcolsep}{3pt}
 \begin{adjustbox}{angle=90}
 \centering
    \begin{tabular}{l|ccc|ccc|ccc|ccc}
    \hline \hline
    & \multicolumn{3}{c}{Band 6 } &       & \multicolumn{1}{l}{Band 7} & \multicolumn{1}{r}{} & \multicolumn{3}{c}{CO(2-1)} & \multicolumn{3}{c}{CO(3-2)} \\ 
    \multicolumn{1}{c}{Source} & \multicolumn{1}{p{5.5em}}{$\theta$\newline{}['']} & \multicolumn{1}{p{4.515em}}{P.A.\newline{}[d]} & \multicolumn{1}{p{5.5em}}{rms\newline{}[mJy/beam]} & $\theta$  & \multicolumn{1}{p{4.5em}}{P.A.\newline{}[d]} & \multicolumn{1}{p{5em}}{rms\newline{}[mJy/beam]} & \multicolumn{1}{p{4.5em}}{Fpeak$^{**}$\newline{}[Jy]} & \multicolumn{1}{p{4.5em}}{$\upsilon_\mathrm{c}^{***}$\newline{}[km/s]} & \multicolumn{1}{p{3.5em}}{$\Delta \upsilon^{****}$\newline{}[km/s]} & \multicolumn{1}{p{5em}}{Fpeak$^{**}$\newline{}[Jy]} & \multicolumn{1}{p{3.5em}}{$\upsilon_\mathrm{c}^{***}$\newline{}[km/s]} & \multicolumn{1}{p{3.5em}}{$\Delta \upsilon^{****}$\newline{}[km/s]} \\
    \hline
    \hline
    \multicolumn{6}{l}{\textit{S-type semi-regulars and irregular stars:}} & \multicolumn{1}{r}{} &       &       & \multicolumn{1}{r}{} &       &       &  \\
    T Cet & 10.5$\times$3.4 & -87.1 & 50  & 4.9$\times$2.8 & 77.2  & 111 & 7.9   &  22.0     &  13.9     & 18.2  &  18.1 & 14.5 \\
    Z Ant & 7.0$\times$4.2 & 82.0  & 47.37 & 6.1$\times$2.5 & 73.6  & 129.5 & 3.5   & -15.9      &  15.8     & 5.5   &  -15.5 & 16.5 \\
    UY Cen & 7.1$\times$4.1 & 81.9  & 65 & 4.5$\times$3.1 & -85.2 & 106.9 & 1.2   & -24.6      & 28.5      & 2.7 &  -24.4 &29.2  \\
    AM Cen & 7.0$\times$6.7 & -49.9 & 69 & 4.4$\times$3.4 & -80.9 & 109 & 4.5   & -27.3      & 8.7      & 7.8   & -27.6& 9.9 \\
    ST Sco & 7.3$\times$4.4 & 69.3  & 40.7 & 4.5$\times$3.0 & 83.5  & 131.5 & 10.8  & -4.4      & 14.0      & 16.0  &-4.3 & 13.8 \\
    FU Mon & 9.1$\times$4.3 & 75.0  & 60  & 5.1$\times$3.0 & -74.6 & 121.6 & 9.8   & -41.5      & 6.7      & 7.5   & -41.8&5.7  \\
    NSV 24833 & 7.7$\times$4.0 & -81.0 & 60.3 & 5.1$\times$2.7 & -88.8 & 86.1 & 4.7   & 57.3      & 20.9  & 6.9 & 58.3 & 21.5 \\
    DY Gem & 6.7$\times$4.7 & -79.2 & 45 & 4.9$\times$3.2 & 67.2  & 137.9 & 4.7   & -16.3      &  21.7     & 8.3   & -16.4&21.3  \\
    RZ Sgr & 6.7$\times$4.4 & 87.4  & 41.59 & 4.6$\times$2.9 & -89.7 & 97.48 & 24.0  & -29.3      &15.9       & 26.1  & -25 & 16.3 \\
    \multicolumn{6}{l}{\textit{S-type Mira stars:}} & \multicolumn{1}{r}{} &       &       &       &       &       &  \\
    TT Cen & 7.5$\times$6.9 & 27.6  & 57.9 & 4.3$\times$3.6 & -80.5 & 128.2 & 8.0   &   6.2    &  43.0     & 13.2  & 6.4 & 42.3 \\
    GI Lup & 6.9$\times$4.2 & 84.0  & 50.1 & 4.8$\times$2.9 & 81.5  & 156.8 & 4.3   &  5.5     &  22.9     & 7.4   &5.3&23.7  \\
    RT Sco & 7.2$\times$4.8 & 67.1  & 46.9 & 4.8$\times$3.1 & 83.8  & 147.1 & 23.3  &  -46.9     &  27.3     & 39.0  & -46.2&28.9  \\
    IRC -10401 & 11.4$\times$3.8 & -82.7 & 53.7 & 5.9$\times$2.9 & -77.7 & 152.9 & 6.0   &  19.4     & 35.2 & 8.9 & 20.5 &37.1  \\
    ST Sgr & 8.4$\times$4.1 & -76.8 & 49.8 & 5.6$\times$2.8 & -84.6 & 81 & 8.0   &  55.9     & 17.0      & 11.6  & 57 & 17.3 \\
    T Sgr & 8.1$\times$4.0 & -79.2 & 44.5 & 5.4$\times$2.7 & -86.2 & 91 & 2.4   & 8.5      & 20.6      & 4.2   & 9.8& 21.2 \\
    W Aql &  -     &    -   &   -    & 5.0$\times$2.6 & 76.9  &    150 &  -     &   -    &   -    &   44.92   &-24.7 & 39.7 \\
    \multicolumn{6}{l}{\textit{M-type semi-regulars ans irregular stars:}} & \multicolumn{1}{r}{} &       &       & \multicolumn{1}{r}{} &       &       &  \\
    $\theta$ Aps & 6.8$\times$6.0 & 48.3  & 45.8 & 4.6$\times$4.1 & 4.3   & 93.4 & 15.5  &    2.6 &  10.2  & 38.9  & 0.8 & 13.1 \\
    R Dor$^*$ & 7.0$\times$5.0 & 78.3  &   53.62    & 4.9$\times$3.5 & -82.2 & 253.5 & 103   & 7.2 & 14.0  & 132.2 &  7.0  & 14.7 \\
    U Men & 8.0$\times$6.4 & 19.8  & 4.6$\times$10\^-2 & 6.9$\times$3.9 & 78.7  &196.5 & 7.0   & 15.1  & 23.8 & 17.1& 15.6 & 25.3 \\
    \multicolumn{6}{l}{\textit{M-type Mira stars:}} & \multicolumn{1}{r}{} &       &       & \multicolumn{1}{r}{} &       &       &  \\
    IRC-30398$^*$ & 9.8$\times$3.7 & -74.1 & 94.2  & 6.7$\times$2.6 & -71.7 & 169.9 & 34.66  & -6.3 & 34.4 & 38.33  & -6.5 & 36.6 \\
    \multicolumn{6}{l}{\textit{C-type semi-regular and irregular stars:}} & \multicolumn{1}{r}{} &       &       & \multicolumn{1}{r}{} &       &       &  \\
    V1302 Cen & 7.1$\times$4.6 & 72.5  & 51 & 5.1$\times$3.3 & 80.0  & 128.5 & 3.5   & -42.4 & 17.8 & 5.8   & -42.8 & 17.8 \\
    V996 Cen & 7.1$\times$4.6 & 75.0  & 48.3 & 4.9$\times$3.2 & 80.5  & 125.6 & 3.5   & -2.0 & 26.3 & 4.8   & -1.9 & 25.6 \\
 X TrA & 6.3$\times$5.7 & 87.0  & 56.84 & 4.3$\times$3.8 & -57.3 & 102.4 & 23.1  & -2.3 & 17.4 & 30.9  & -3.8 & 20.0 \\
    AQ Sgr &   -    &   -    &   -    & 4.5$\times$2.7 & -85.6 & 136.4 &   - & - &    -   & 12.9  &  21.6     &  23.8\\
    \multicolumn{6}{l}{\textit{C-type Mira stars:}} & \multicolumn{1}{r}{} &       &       & \multicolumn{1}{r}{} &       &       &  \\
    V821 Her & 7.2$\times$5.9 & -87.5 & 49.3 & 4.3$\times$3.6 & -59.6 & 124.6 & 48.4  & -0.3 & 27.2& 57.1  &-0.2  & 27.1 \\
    AI Vol$^*$ & 6.4$\times$5.5 & 89.8  & 44.23 & 4.5$\times$3.7 & -87.3 & 163.1 & 78.5  & -38.9 & 26.6 & 75.88 & -38.8 &26.2\\
    AFGL 3068$^*$ & 7.2$\times$5.4 & 66.9  & 100.6 & 4.9$\times$3.6 & 8.1   & 185.3 & 70.67  & -31  & 29& 81.26  & -31.4  & 28.7 \\
    \bottomrule
    \multicolumn{8}{l}{\footnotesize  $\theta$ is the beam size, P.A. the position angle, $F_\mathrm{peak}$ the peak flux, $\upsilon_\mathrm{c}$ the centre velocity, and $\Delta \upsilon$ the velocity width} \\
    \multicolumn{8}{l}{\footnotesize  *Observed with TP   } \\
     \multicolumn{8}{l}{\footnotesize  **The peak flux is the maximum point across the line   } \\
     \multicolumn{8}{l}{\footnotesize  ***The centre velocity is the average of the two points at the extreme velocities that correspond to five percent peak flux } \\
     \multicolumn{8}{l}{\footnotesize  ****The velocity width between the two five-percent-peak-flux points gives the total width. } \\
    \end{tabular}%
  \end{adjustbox}
  \label{tab:imaging_results}%
\end{table*}%

\begin{figure*}
\centering
   \includegraphics[width=1.07\textwidth]{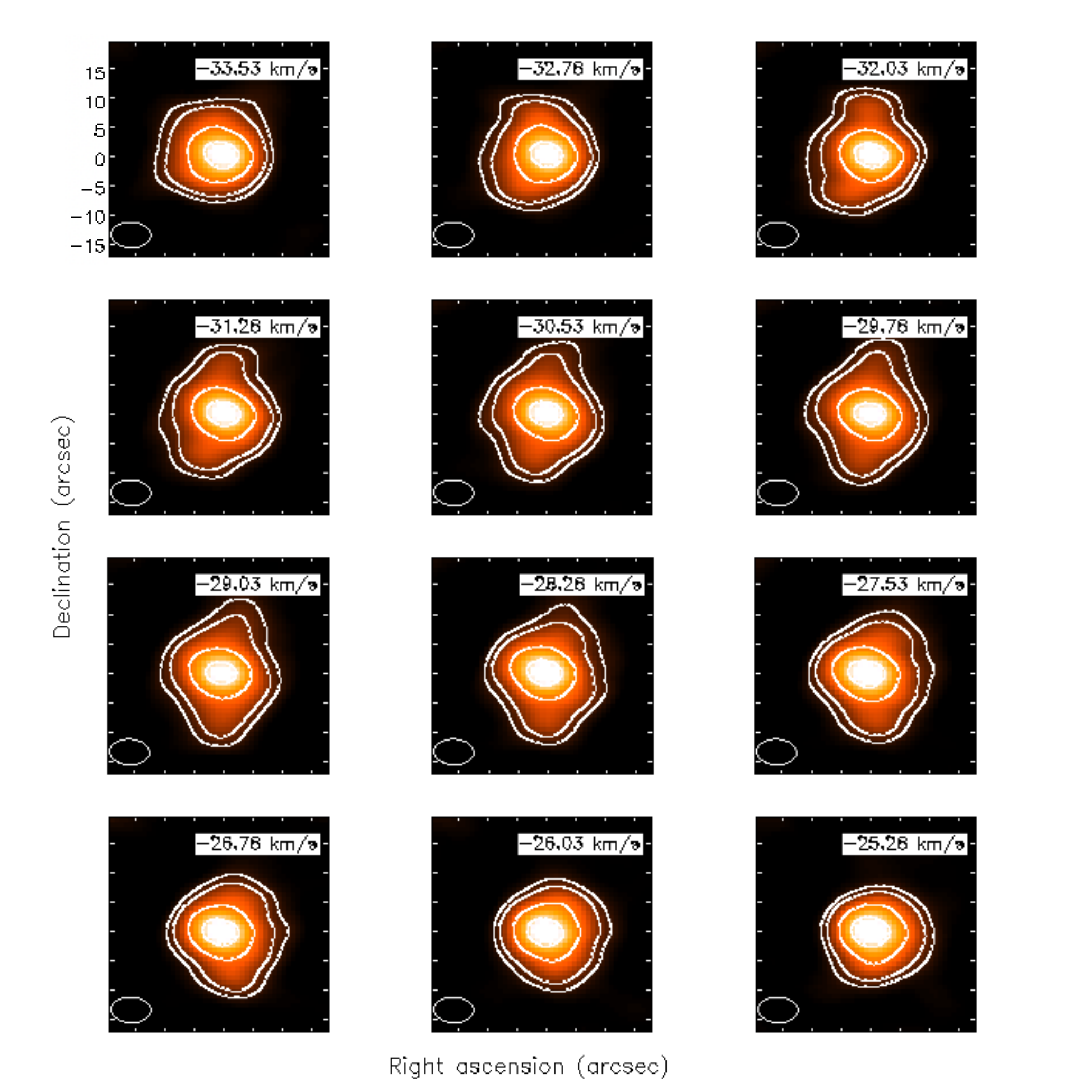}
    \caption{Channel maps of RZ Sgr in the CO(2--1) line. The contour levels correspond to 20, 40, 60, and 100 times the rms level. The ellipse in the bottom-left corner of each image represents the beam. }
    \label{fig:RZ_Sgr_channel21}
\end{figure*}
\begin{figure*}
\centering
   \includegraphics[width=1.07\textwidth]{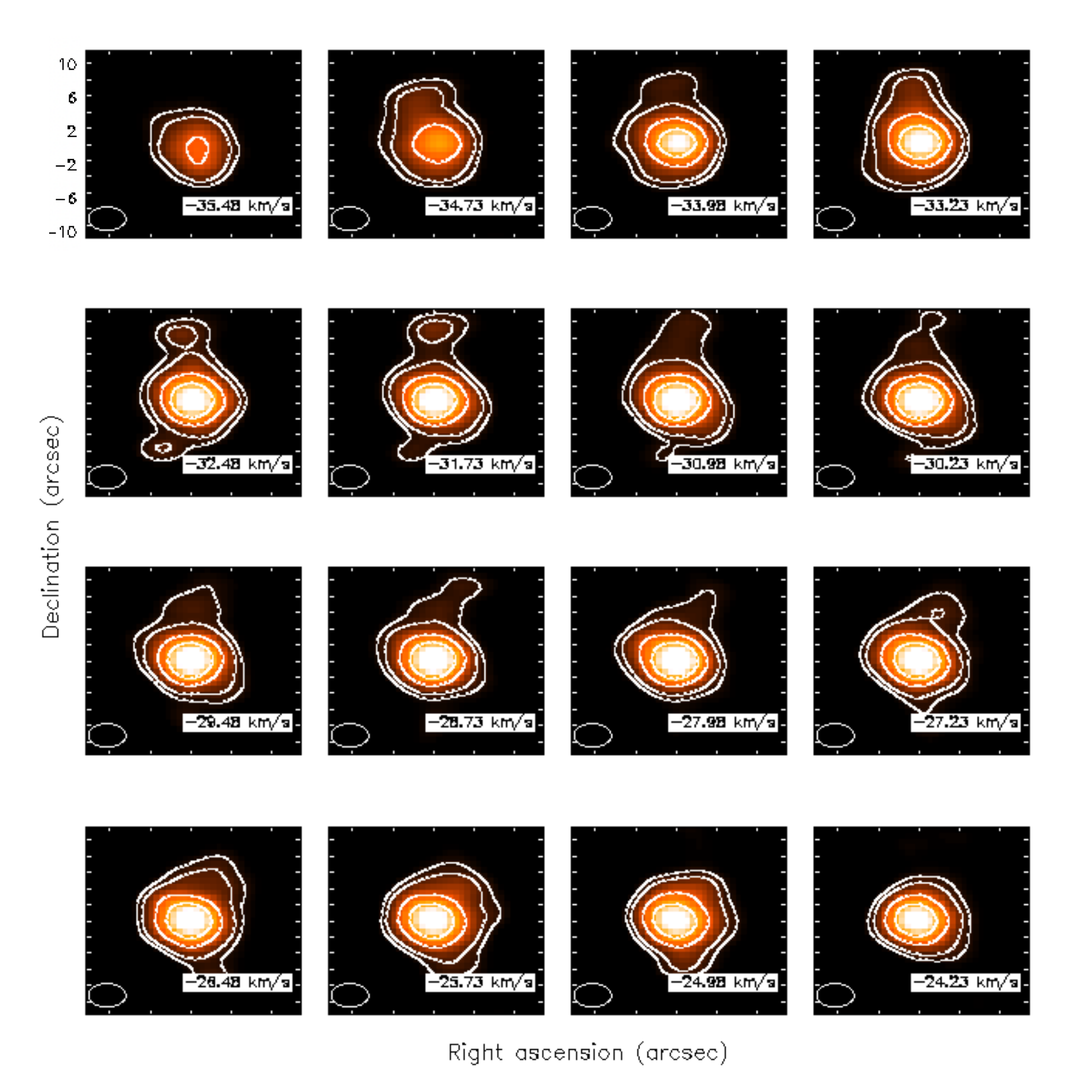}
    \caption{Channel maps of RZ Sgr in the CO(3--2) line. The contour levels correspond to 5, 10, 30, and 60 times the rms level. The ellipse in the bottom-left corner of each image represents the beam. }
    \label{fig:RZ_Sgr_channel32}
\end{figure*}
\begin{figure*}
\centering
   \includegraphics[width=0.9\textwidth]{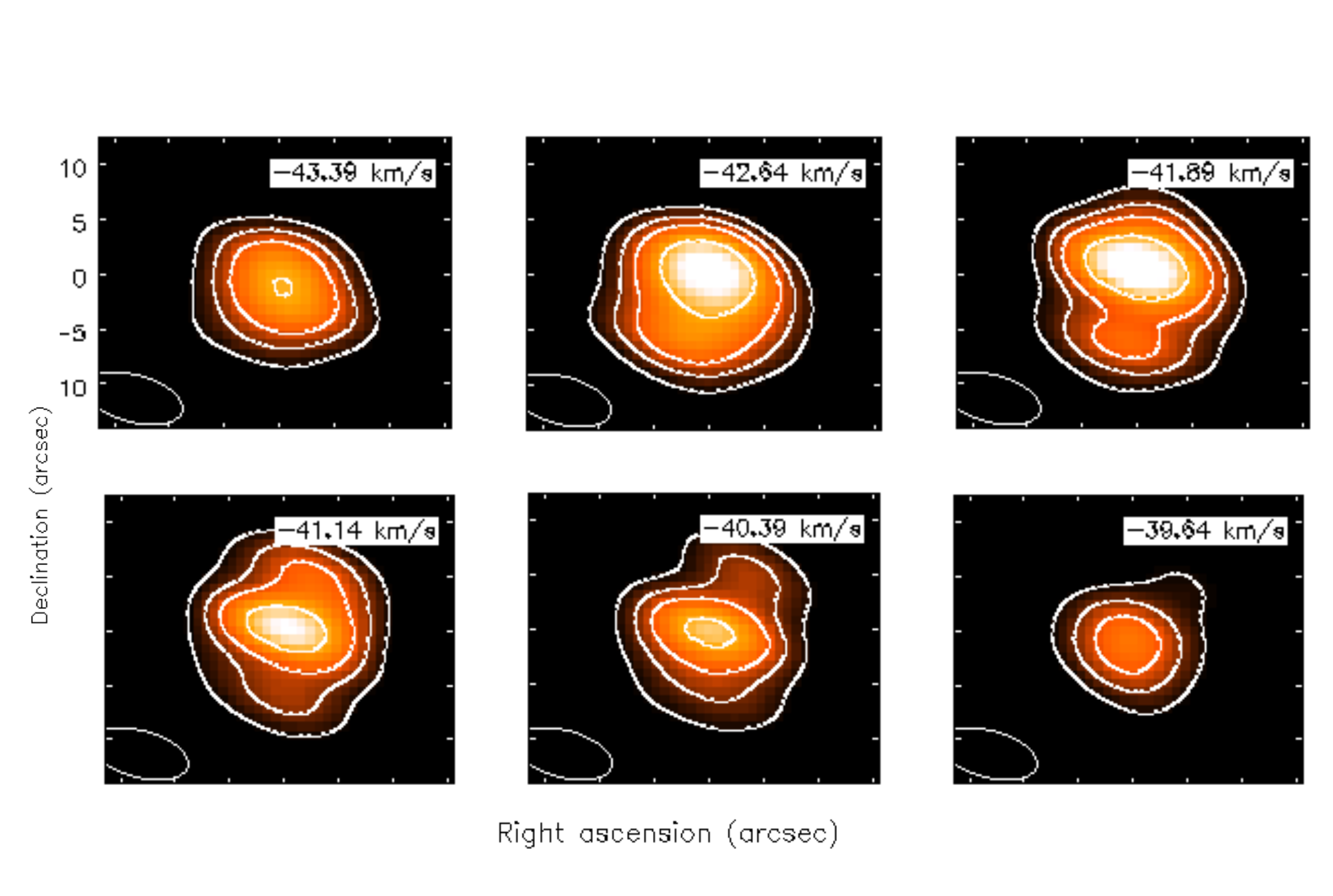}
    \caption{Channel maps of FU Mon in the CO(2--1) line. The contour levels correspond to 7, 14, 22, and 40 times the rms level. The ellipse in the bottom-left corner of each image represents the beam. }
    \label{fig:FU_Mon_channel21}
\end{figure*}
\begin{figure*}
\centering
   \includegraphics[width=0.9\textwidth]{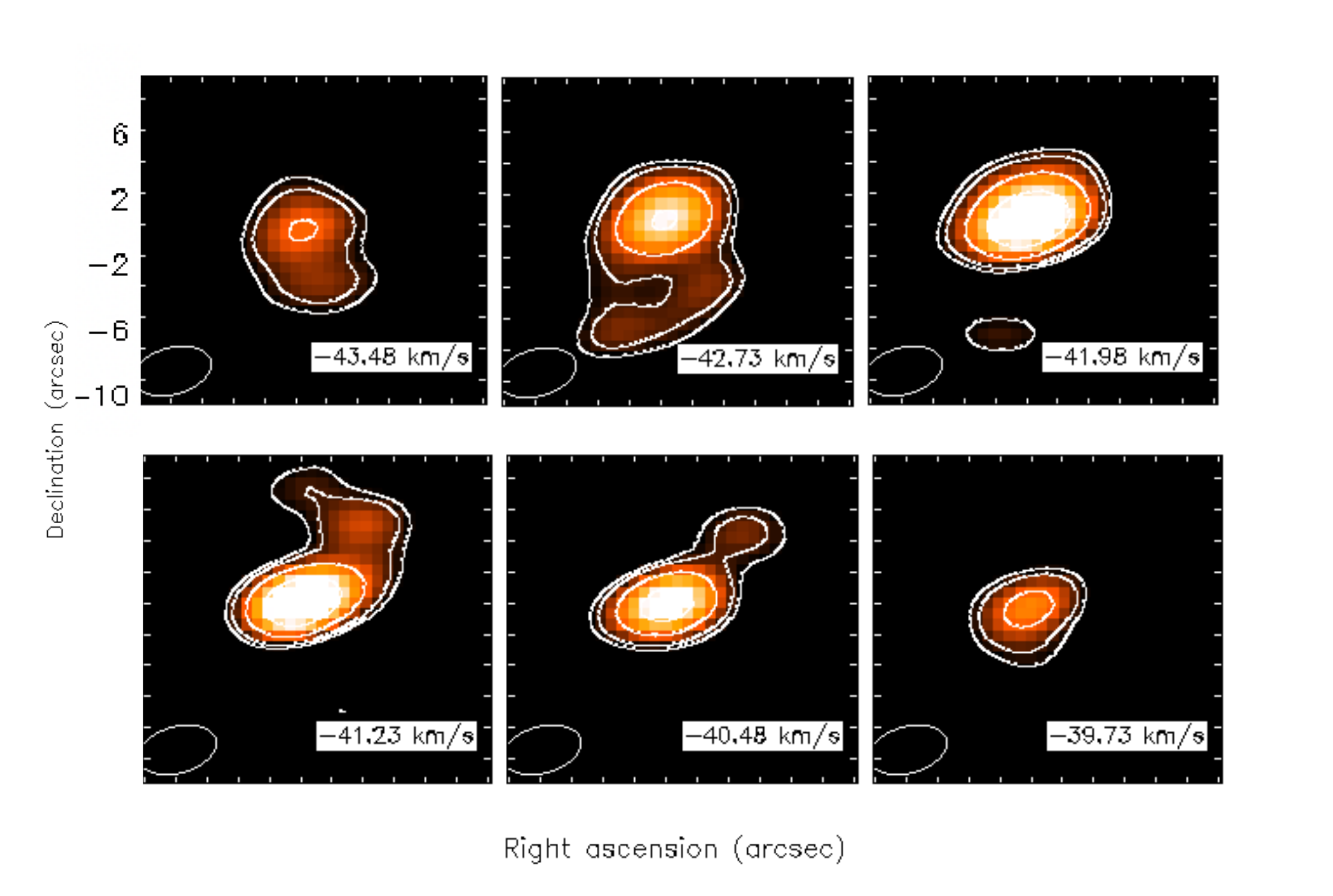}
    \caption{Channel maps of FU Mon in the CO(3--2) line. The contour levels correspond to 3.5, 5, 10, and 20 times the rms level. The ellipse in the bottom-left corner of each image represents the beam. }
    \label{fig:FU_Mon_channel32}
\end{figure*}
\begin{figure*}
\centering
   \includegraphics[width=1.07\textwidth]{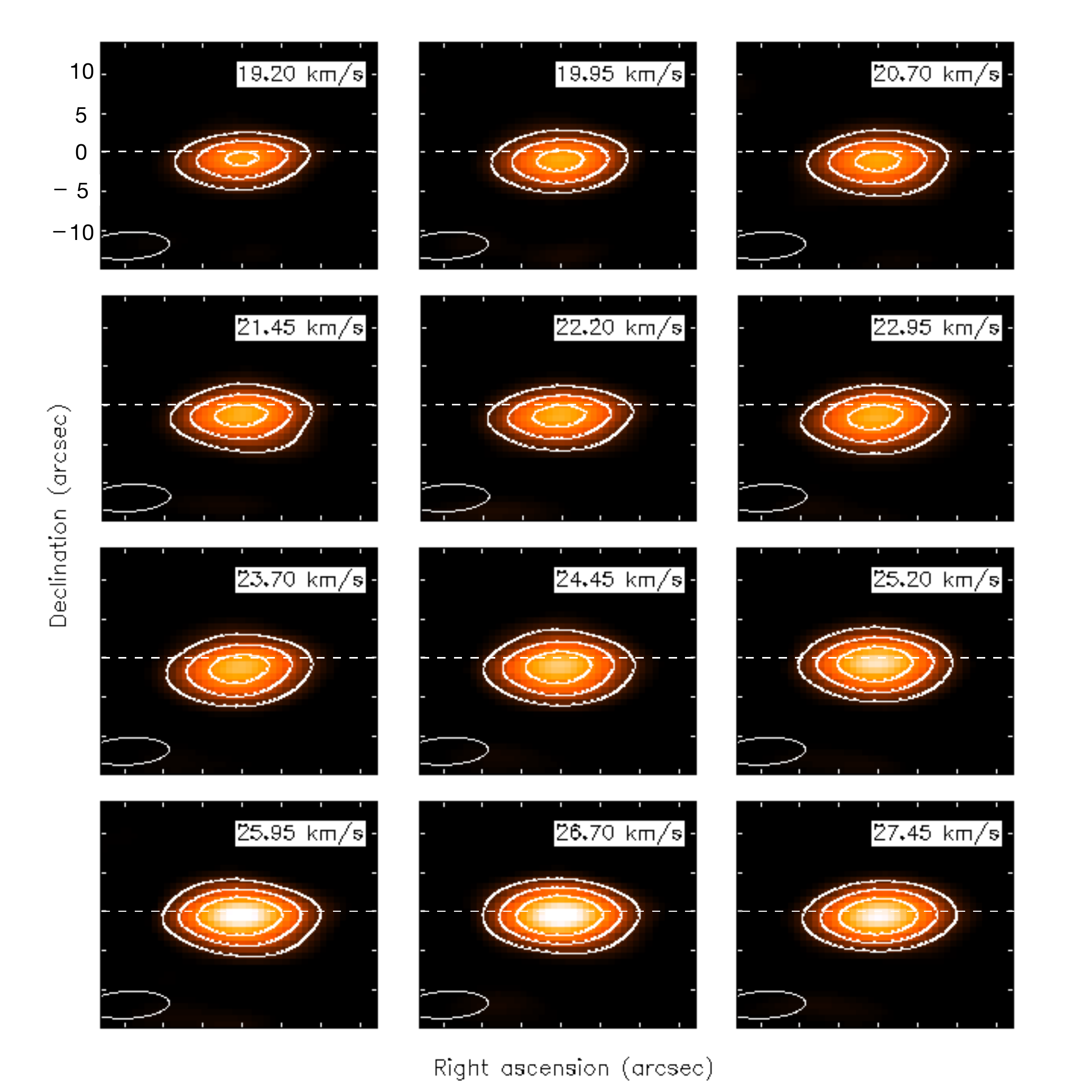}
    \caption{Channel maps of T Cet in the CO(2--1) line. The contour levels correspond to 10, 20. and 50 times the rms level. The ellipse in the bottom-left corner of each image represents the beam. The white dashed line cuts each image in half at a relative declination of 0 arcsec.}
    \label{fig:T_Cet_channel21}
\end{figure*}
\begin{figure*}
\centering
   \includegraphics[width=1.07\textwidth]{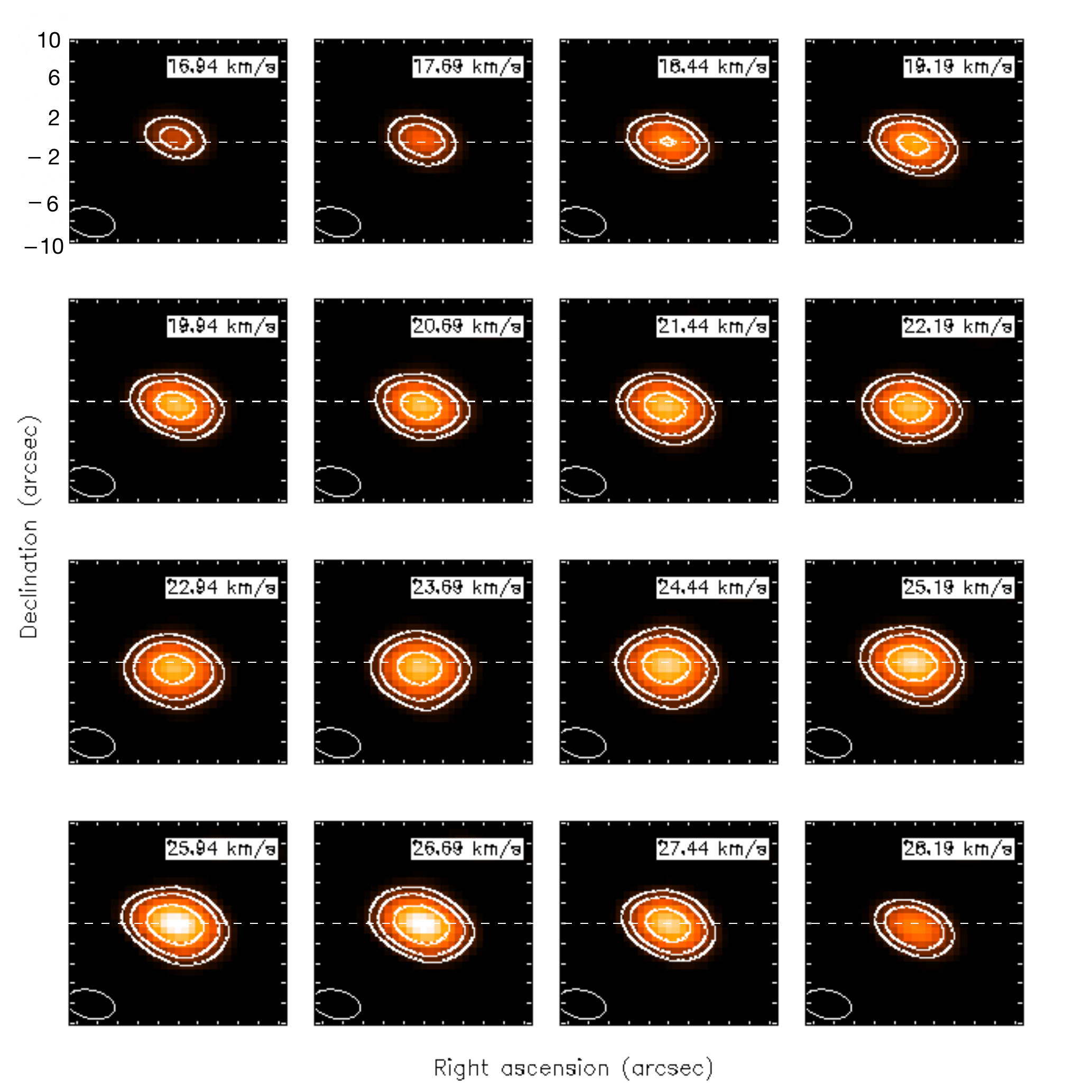}
    \caption{Channel maps of T Cet in the CO(3--2) line. The contour levels correspond to 10, 20, and 50 times the rms level. The ellipse in the bottom-left corner of each image represents the beam. The white dashed line cuts each image in half at a relative declination of 0 arcsec.}
    \label{fig:T_Cet_channel32}
\end{figure*}
\section{Detection of molecules other than $^{12}$CO}
\begin{table*}[!htbp]
\centering
  \caption{Peak flux of detected molecular emission measured within a circular 10$^{\prime \prime}$ aperture centred on the S-type stars.} 
\setlength{\tabcolsep}{5pt}
 \begin{adjustbox}{angle=360}
  \begin{tabular}{lccccc} 
      \toprule
    \hline
    \multirow{3}{*}{Source $\backslash$ Frequency$^*$ [GHz]} &  \multicolumn{5}{c}{Peak Flux [Jy]} \\

 & SiO ($\nu =1$, 5--4)  & SiO ($\nu =0$, 5--4) &  $^{13}$CO (3--2)  & CS (7--6) &  $^{29}$SiO (8--7) \\
     & 215.596 & 217.105 & 330.588 & 342.883 & 342.980\\
    \midrule
    T Cet & - & 0.2 & 1 & - & 0.4\\
    Z Ant & - & 0.6& 0.3 &- & 0.4 \\
    UY Cen & - & - & -&-&-\\
    AM Cen & - & -& 0.4& - & -\\
    ST Sco & - & 0.8 &0.8 &-& 0.8\\
    FU Mon & - & - & 0.5 & -& -\\
    NSV 24833 & 0.6 & 0.8 & 1.4& - &0.3\\
    DY Gem & - & 0.2 & 0.4 & - & 0.1 \\
    RZ Sgr & 0.1 &1.1& 5.4&0.08&0.3\\
    TT Cen & - & 0.1& 0.8 & - &- \\
    GI Lup & - & 0.3 & 0.4 & 0.2 & 0.2 \\
    RT Sco & 0.8 & 2.3 & 0.3 & 0.5 & 1.8 \\
    IRC-10401 & 3 & 1.5 & 0.4 & 0.2 & 0.5\\
    ST Sgr & 0.1 & 1& 0.8 & 0.1 & 0.7 \\
    T Sgr& 0.2 &0.1& 0.2& - & 0.1\\

    \bottomrule
  \multicolumn{4}{l}{\footnotesize *The line frequencies are taken from Splatalogue.} \\
  \end{tabular}
  \end{adjustbox}
  \label{tab:other_molecules}%
\end{table*}%

\section{Line profiles}
\begin{figure*}
  \centering
  \includegraphics[page=1,width=0.9\textwidth]{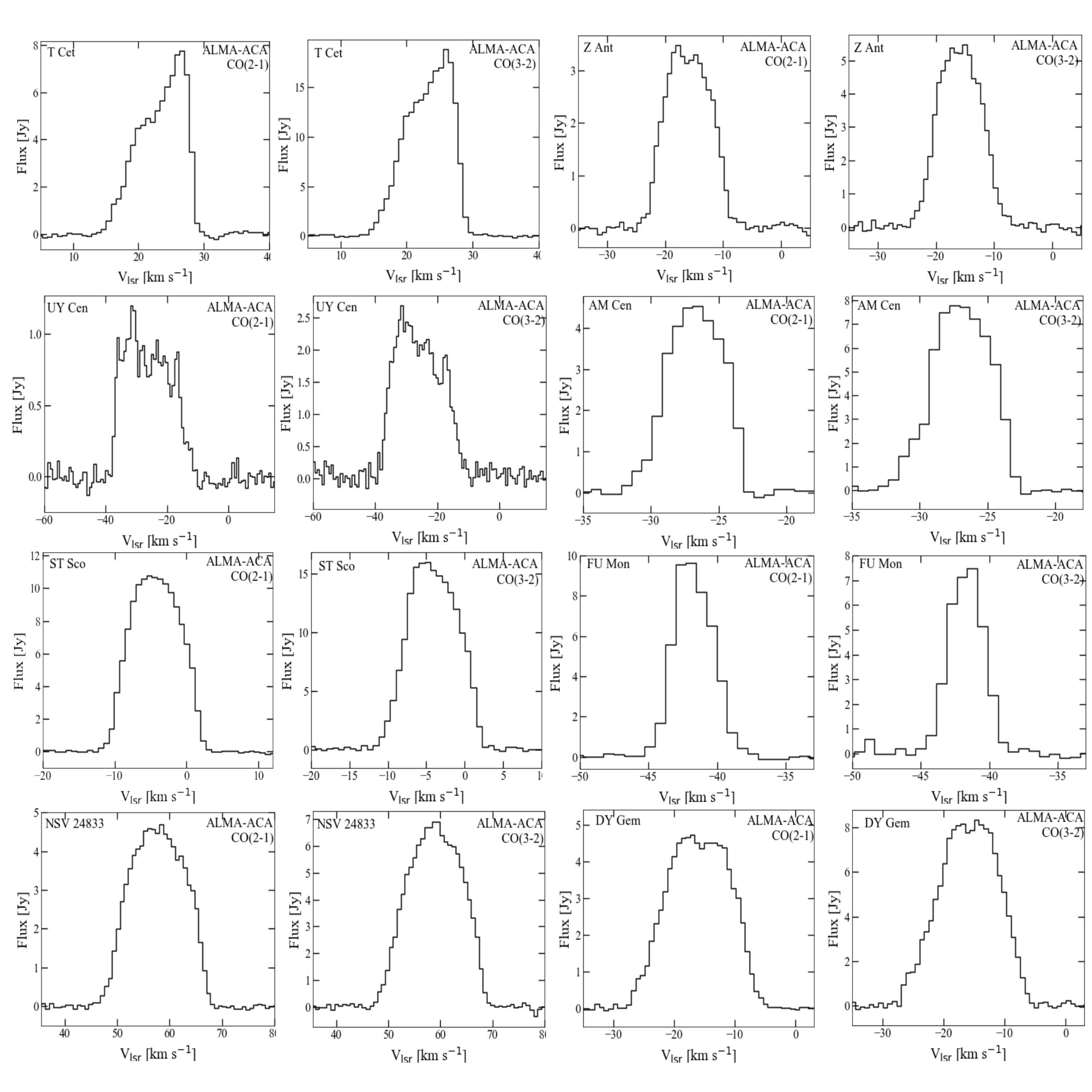}
    \caption{CO J = 2--1 and 3--2 line profiles measured towards the S-type AGB stars of the sample discussed in this paper. The source name is given in the upper right corner and the transition is in the upper left corner of each plot.}
    \label{fig:lines_profiles_S1}
\end{figure*}
\begin{figure*}
   \includegraphics[page=2,width=0.9\textwidth]{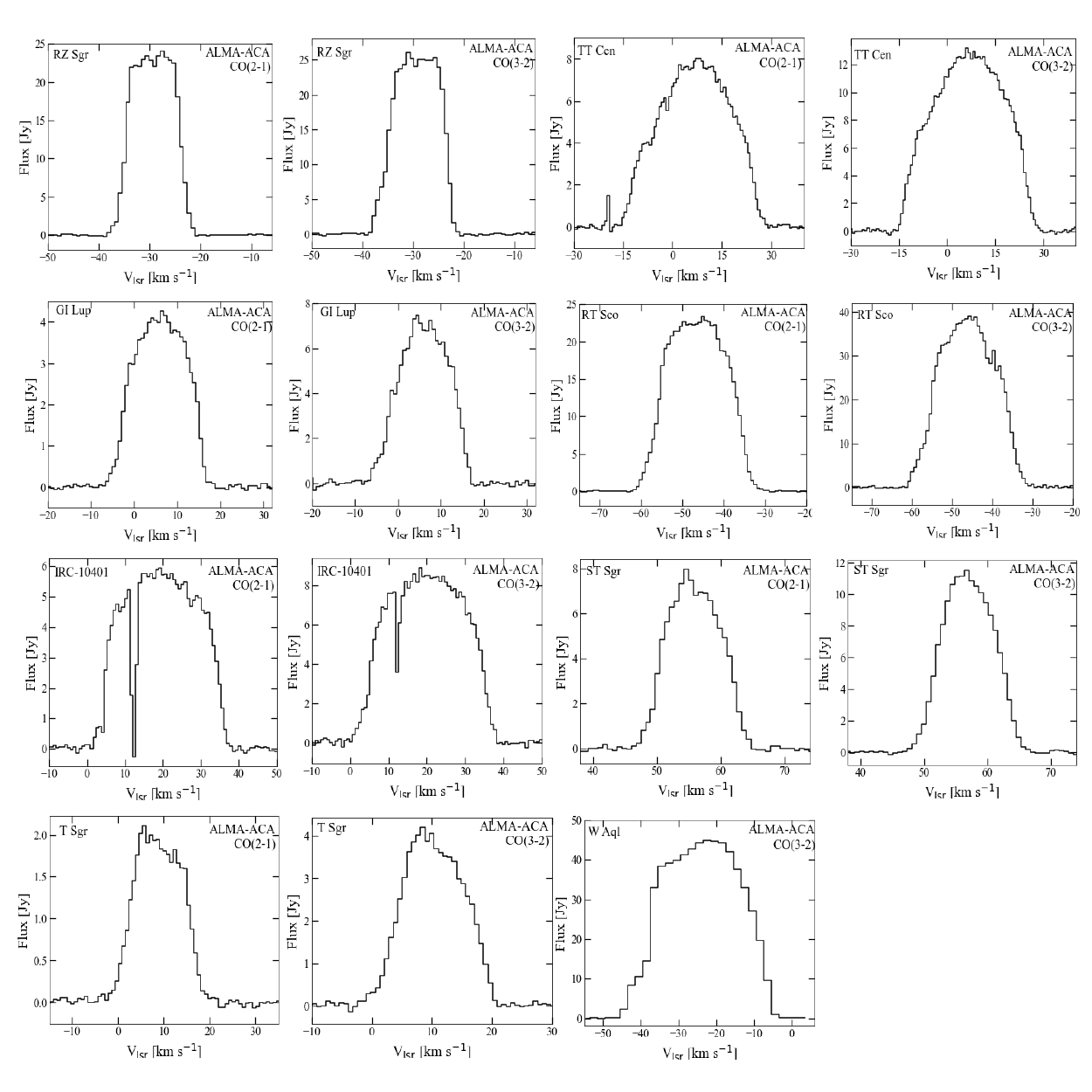}
    \caption{CO J = 2--1 and 3--2 line profiles measured towards the S-type AGB stars of the sample discussed in this paper. The source name is given in the upper right corner and the transition is in the upper left corner of each plot.}
    \label{fig:lines_profiles_S2}
\end{figure*}
\begin{figure*}
   \includegraphics[page=3,width=0.9\textwidth]{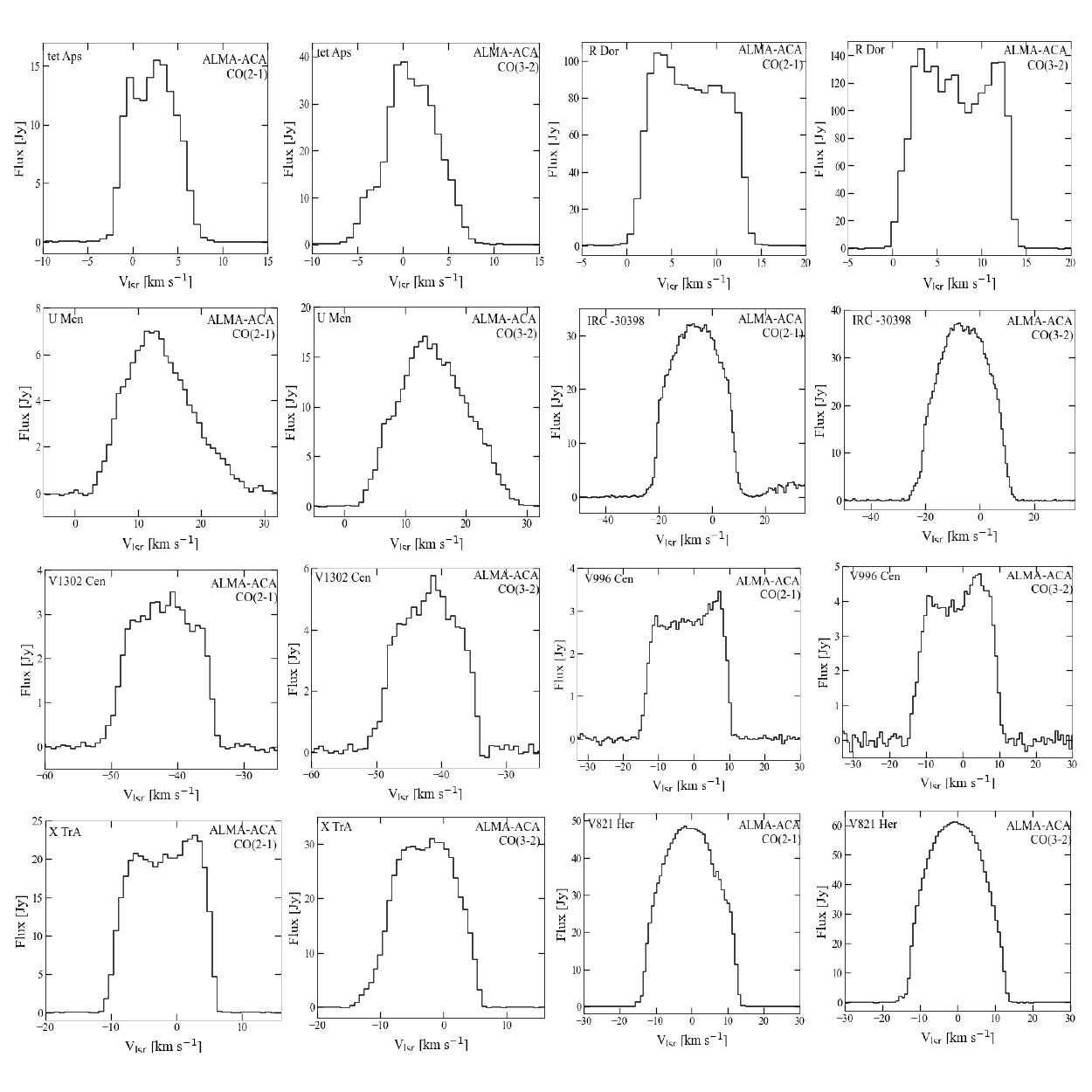}
    \caption{CO J = 2--1 and 3--2 line profiles measured towards the M- and C-type AGB stars of the sample discussed in this paper. The source name is given in the upper right corner and the transition is in the upper left corner of each plot.}
    \label{fig:lines_profiles_S3}
\end{figure*}
\begin{figure*}
   \includegraphics[width=0.85\textwidth]{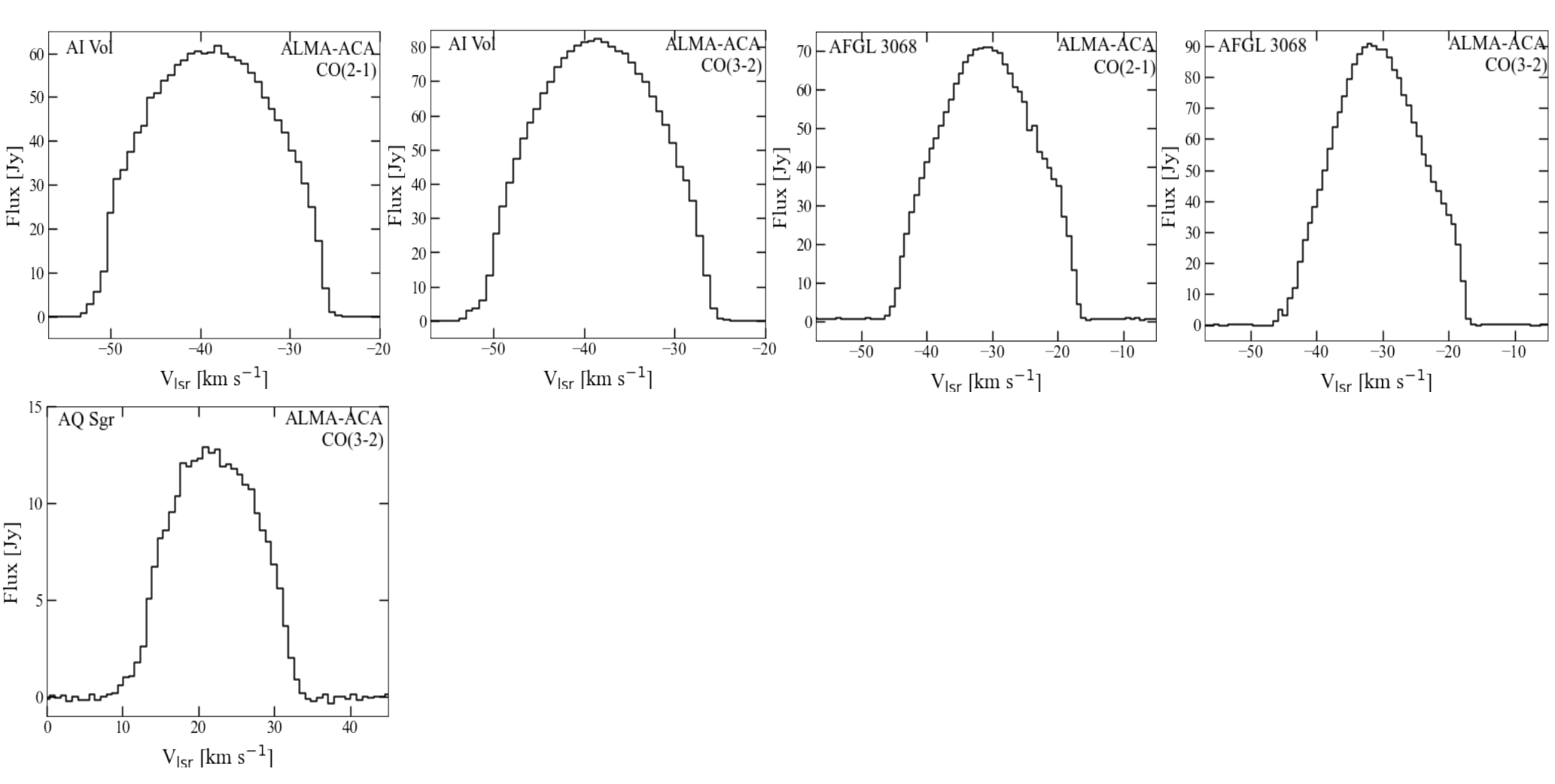}
    \caption{CO J = 2--1 and 3--2 line profiles measured towards the M- and C-type AGB stars of the sample discussed in this paper. The source name is given in the upper right corner and the transition is in the upper left corner of each plot.}
    \label{fig:lines_profiles_S4}
\end{figure*}
\section{Results from fitting to Gaussian emission distribution}
\begin{figure*}
  \centering
  \includegraphics[page=1,width=0.88\textwidth]{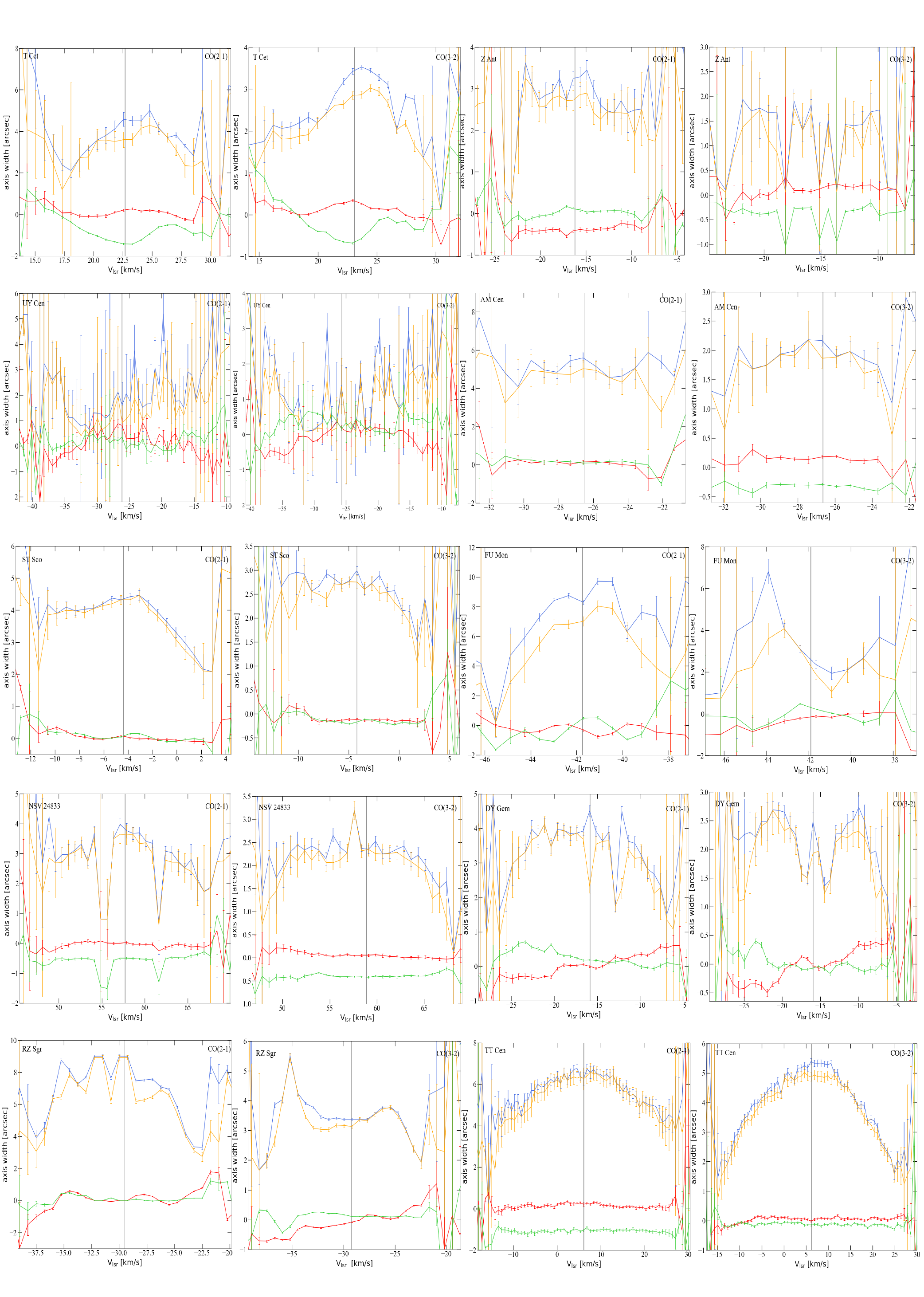}
    \caption{Results from the visibility fitting to the data measured towards the S-type AGB stars of the sample discussed in this paper. The source name is given in the upper left corner and the transition is in the upper right corner of each plot. The upper blue and orange lines show the major and minor axis of the best-fitting Gaussian in each channel, respectively. The lower red and green lines show the RA and Dec offset relative to the centre position, respectively.}
    \label{fig:UV_fit_S1}
\end{figure*}
\begin{figure*}
   \includegraphics[width=0.88\textwidth]{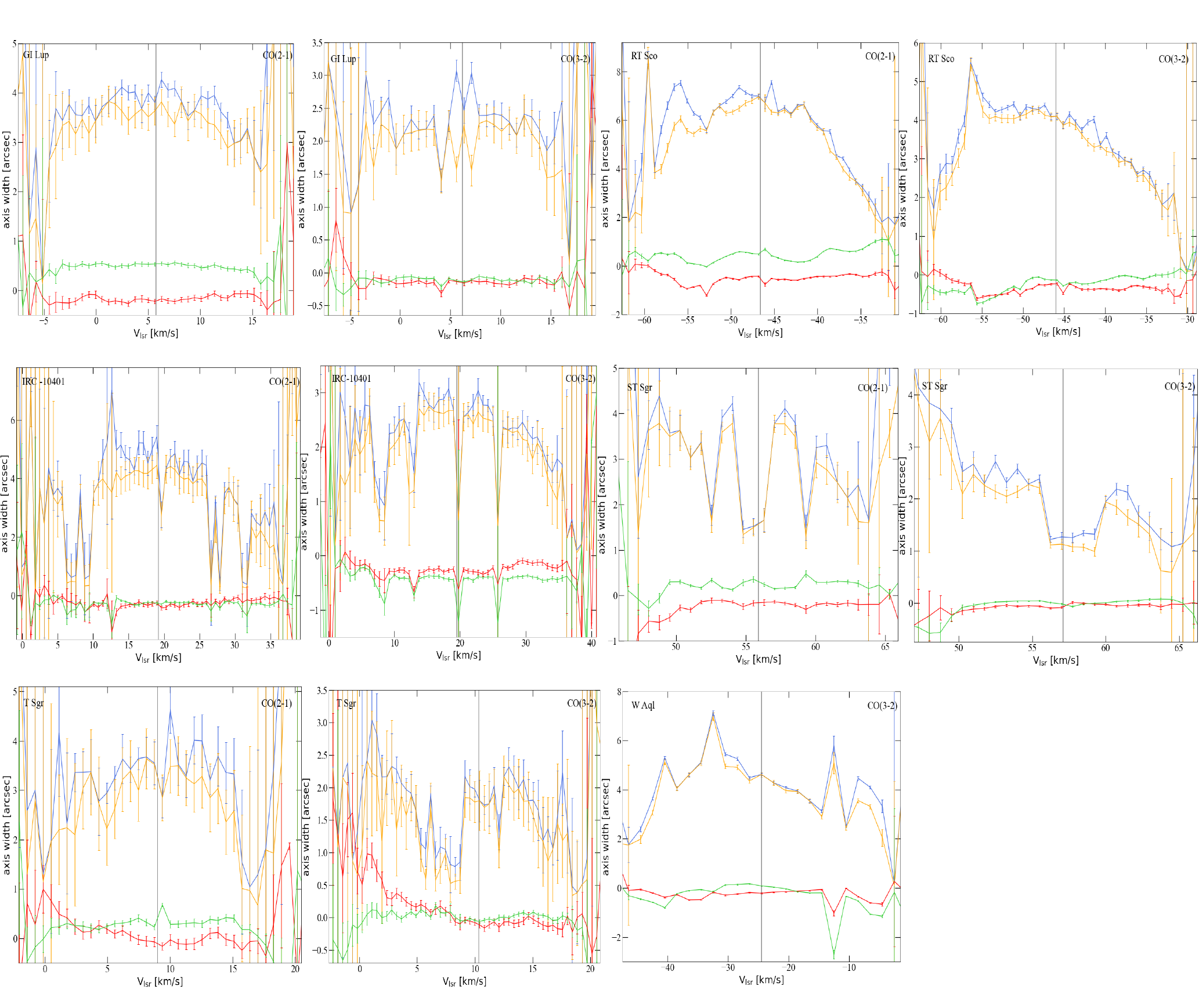}
    \caption{Results from the visibility fitting to the data measured towards the S-type AGB stars of the sample discussed in this paper. The source name is given in the upper left corner and the transition is in the upper right corner of each plot. The upper blue and orange lines show the major and minor axis of the best-fitting Gaussian in each channel, respectively. The lower red and green lines show the RA and Dec offset relative to the centre position, respectively.}
    \label{fig:UV_fit_S2}
\end{figure*}
\begin{figure*}
   \includegraphics[page=3,width=0.88\textwidth]{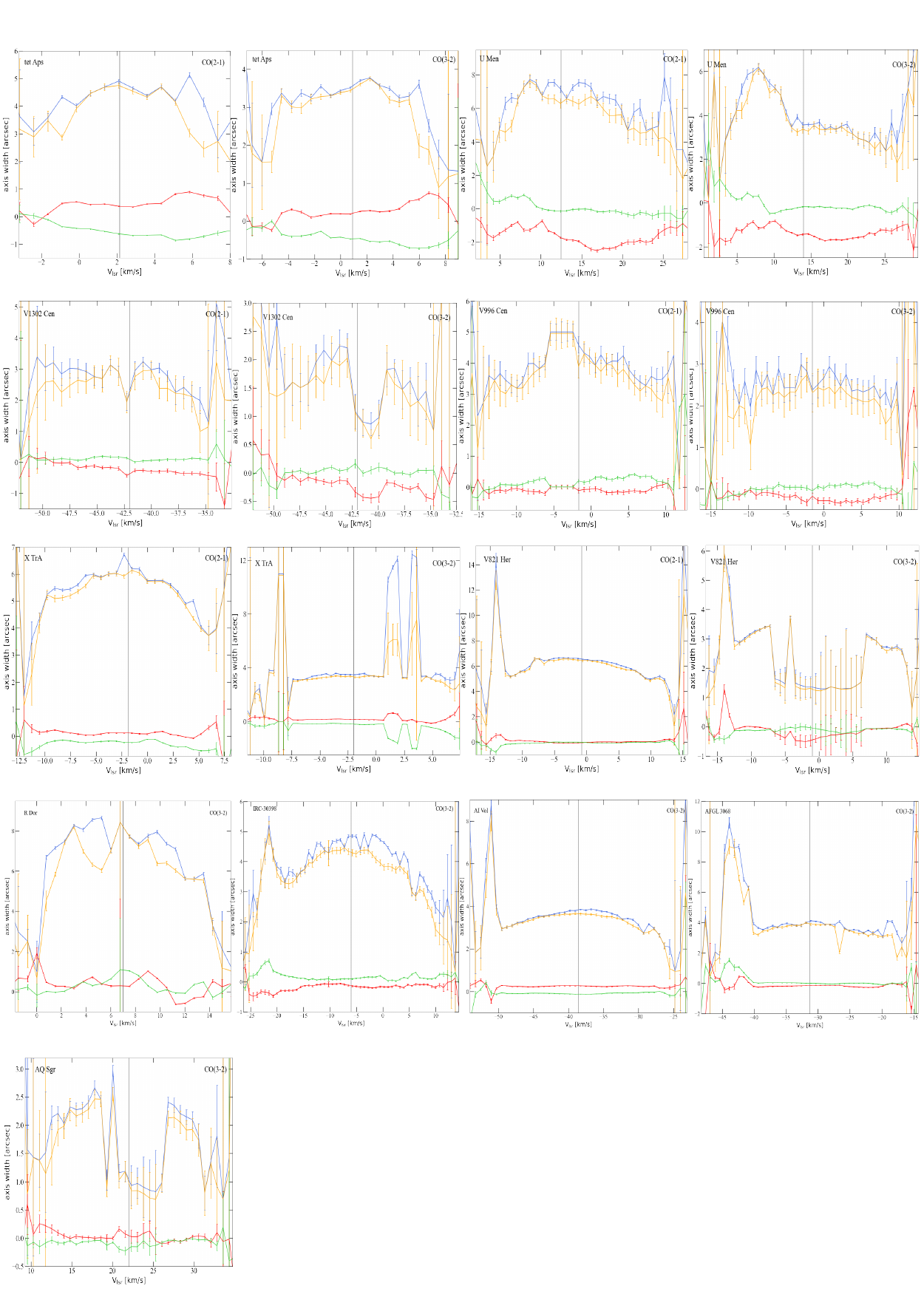}
    \caption{Results from the visibility fitting to the data measured towards the M- and C-type AGB stars of the sample discussed in this paper. The source name is given in the upper left corner and the transition is in the upper right corner of each plot. The upper blue and orange lines show the major and minor axis of the best-fitting Gaussian in each channel, respectively. The lower red and green lines show the RA and Dec offset relative to the centre position, respectively.}
    \label{fig:UV_fit_MC}
\end{figure*}

\end{appendix}
\end{document}